\RequirePackage{ifpdf}
\ifpdf % We are running pdfTeX in pdf mode
\documentclass[pdftex]{sigma}
\else
\documentclass{sigma}
\fi

\numberwithin{equation}{section}
\numberwithin{theorem}{section}
\numberwithin{proposition}{section}
\numberwithin{lemma}{section}
\numberwithin{corollary}{section}
\numberwithin{definition}{section}
\numberwithin{example}{section}
\numberwithin{remark}{section}
\begin{document}

\allowdisplaybreaks

\renewcommand{\PaperNumber}{073}

\FirstPageHeading

\ShortArticleName{On Linear Dif\/ferential Equations Involving a Para-Grassmann Variable}

\ArticleName{On Linear Dif\/ferential Equations\\ Involving a Para-Grassmann Variable}

\Author{Touf\/ik MANSOUR~$^\dag$ and Matthias SCHORK~$^\ddag$}

\AuthorNameForHeading{T. Mansour and M. Schork}

\Address{$^\dag$~Department of Mathematics, University of Haifa, 31905 Haifa, Israel}
\EmailD{\href{mailto:toufik@math.haifa.ac.il}{toufik@math.haifa.ac.il}}
\URLaddressD{\url{http://www.math.haifa.ac.il/toufik/}}

\Address{$^\ddag$~Camillo-Sitte-Weg 25, 60488 Frankfurt, Germany}
\EmailD{\href{mailto:mschork@member.ams.org}{mschork@member.ams.org}}

\ArticleDates{Received May 01, 2009, in f\/inal form July 05, 2009;  Published online July 15, 2009}

\Abstract{As a f\/irst step towards a theory of dif\/ferential equations
involving para-Grassmann variables the linear equations with
constant coef\/f\/icients are discussed and solutions for equations of
low order are given explicitly. A connection to $n$-generalized Fibonacci
numbers is established. Several other classes of dif\/ferential equations (systems of f\/irst order, equations with variable coef\/f\/icients, nonlinear equations) are also considered and the analogies or dif\/ferences to the usual (``bosonic'') dif\/ferential equations discussed.}


\Classification{11B39; 13A99; 15A75; 34A30; 81R05; 81T60}

\section{Introduction}
\def\TT{{\bf T}}
\def\FF{{\bf F}}
Schwinger introduced anticommuting variables (i.e., Grassmann algebras) into physics in order to extend his treatment to fermions of quantum f\/ields using Green's functions and sources~\cite{Sch}. A few year later Martin extended Feynman's path integral method of quantization to systems containing fermions and needed a ``classical'' fermion to quantize, thereby extending Schwinger's analysis of Grassmann algebras~\cite{Mar}. These anticommuting variables $\theta_i$ satisfying $\theta_i\theta_j=-\theta_j\theta_i$~-- implying in particular that $\theta_i^2=0$ -- were used by a number of authors to develop a systema\-tic quantization of fermions in analogy to the well-known quantization of bosons based on the conventional commuting variables, see, e.g., the monograph of Berezin \cite{Ber} for an early survey. Around 1970 the idea of a symmetry between bosons and fermions -- called supersymmetry -- emerged (for accounts on the early history see, e.g.,~\cite{KS}), providing a link between particles having dif\/ferent statistics. It took then only a few years until the commuting and anticommuting variables were merged and interpreted as coordinates on a ``superspace'' (mainly due to Akulov and Volkov) and the physical superf\/ields as ``functions'' on superspace (mainly due to Salam, Strathdee, Ferrara, Wess and Zumino). This happened in 1974 and from then on the idea of a physical and  mathematical superworld has captured the imagination of physicists and mathematicians alike (see, e.g., \cite{WB,Fre} for a physical discussion and \cite{Wit,Man,Rog} for a mathematical discussion). In our context it is particularly interesting to note that the introduction of the anticommuting (and nilpotent) variables has led to beautiful extensions of well-known mathematical structures (supermanifolds, supergroups, etc.).

In a rather dif\/ferent line of thought Green introduced in 1953 parastatistics into quantum f\/ield theory by considering certain trilinear commutation relations, generalizing the quantization schemes for bosons and fermions \cite{Gre}. With hindsight it is not surprising what the next steps should have been in analogy to Fermi--Dirac statistics (i.e., fermions): The introduction of the ``classical paraobjects'' which yield upon quantization the operators satisfying the trilinear commutation relations (and are nilpotent in higher order). This should motivate the introduction of para-Grassmann algebras and then, in close analogy to above, a merging of commuting and para-Grassmann variables as coordinates on a parasuperspace and describing physical parasuperf\/ields as functions on this ``space''. From a more mathematical point of view one would hope that these parastructures would again yield fascinating new extensions of well-known structures. In reality the development took some time; although there had been precursors \cite{Mar,Kal} the real development took of\/f around 1979 with the work of Kamefuchi and coworkes \cite{OmK1,OmK2,OhK1}. In their work the focus was on the quantum f\/ield theory of particles satisfying parastatistics, see, e.g., the early monograph \cite{OhK2}. Para-Grassmann variables -- satisfying $[\theta_i, [\theta_j,\theta_k]]=0$ as well as $[\theta_{i_1},\theta_{i_2},\ldots,\theta_{i_m}]_{+}=0$ where $m\geq p+1$ and $[.,\ldots,.]_+$ denotes the completely symmetrized pro\-duct (note that this implies $\theta_i^{p+1}=0$) - were mainly considered as a convenient tool, but already in~\cite{OmK2} parasuperf\/ields depending on conventional and para-Grassmann variables were introduced in close analogy to superf\/ields. One of the f\/irst applications were in a model of one-dimensional supergravity interacting with ``matter'' represented by para-Grassmann-variables \cite{GT} and in a~para-Grassmann extension of the Neveu--Schwarz--Ramond algebra \cite{Zhe}. Further applications have been in a parasupersymmetric extension of quantum mechanics \cite{RuS,Yam1,DFMV,DMSV,FRY}, in  particular in generalizations of the Pauli and Dirac equations \cite{Yam2,Yam3} to describe particles with spin $s>\frac{1}{2}$, as well as in the closely related fractional supersymmetry \cite{Dur1,Dur2,Dur3,AM,DMAP}. Low dimensional f\/ield theories, in particular conformal f\/ield theories, have also been discussed \cite{ABC,SSZ1,SSZ2,FR,PRS,RS,KK1,KK2,KK3,SZ} and a generalization of the Virasoro algebra has been disco\-ve\-red \cite{Dur4,FIK1}. In particular in this approach the geometrical point of view of merging the ordinary variables with the para-Grassmann variables into a set of coordinates on a parasuperspace and considering the fractional supersymmetric f\/ields as functions on this space has been used in close analogy to the well-known superspace and superf\/ields. From a dif\/ferent point of view a~connection to deformed algebras or quantum groups was reported \cite{FIK2,rauschcliff,ABCD,Isa,Ply1,Ply2,Alv,CMT}. Of course, para-Grassmann variables have also been considered in their own right \cite{FRausch,FIK3,CLMS,matthias}. Many more references can be found in \cite{Rauschsurv} which surveys also some of these developments.

Let us point out that the para-Grassmann variables considered originally (satisfying trilinear relations) lead quickly to rather messy structures and calculations. Therefore, a generalization of the Grassmann variables was introduced by Filippov, Isaev and Kurdikov in \cite{FIK1,FIK2,FIK3} satisfying $\theta_i^{p+1}=0$ as well as certain bilinear commutation relations of the form $\theta_i\theta_i=e^{\frac{2\pi i}{p+1}}\theta_j\theta_i$ (for $i<j$). In most of the more recent works (i.e., after 1992) these generalized Grassmann variables~-- also called para-Grassmann variables -- are used. Of course, in the single-variable case these dif\/ferent approaches reduce to the same sole condition $\theta^{p+1}=0$ and the correspon\-ding structure including a parasupercovariant derivative has been introduced in \cite{ABC} (a formal variable~$\lambda$ satisfying \mbox{$\lambda^{p+1}=0$} was already considered by Martin in \cite{Mar}).

From a more mathematical point of view one may hope that there exists an associated
$\mathbf{Z}_{p+1}$-graded mathematics, yielding for $p=1$ ``ordinary''
supermathematics. Some results in this direction have already been
obtained in the literature cited above. In particular, in the paper~\cite{matthias} by one of the authors many basic elements of para-Grassmann calculus (involving a~single para-Grassmann variable) are collected and some simple observations concerning dif\/ferential equations involving a para-Grassmann variable were obtained. It is the aim of the present paper to start a~systematic treatment of dif\/ferential equations involving para-Grassmann variables.

Before turning to a more detailed description of the present paper we would like to point out that in \cite{Alv} a dif\/ferential equation containing a para-Grassmann variable is solved in connection with deformed coherent states. To the best of our knowledge this is the only paper where such a dif\/ferential equation is solved explicitly (more details can be found below).

Now, let us describe the contents of the present paper in more detail. In
Section~\ref{section2} the def\/initions and basic properties of para-Grassmann
calculus are recalled and the notation is f\/ixed. In Section~\ref{section3} some of
the simplest dif\/ferential equations are solved and compared to the usual ``bosonic'' dif\/ferential equations. The linear dif\/ferential equations of
second order (with constant coef\/f\/icients) are considered in Section~\ref{section4}. In the direct approach chosen a recurrence relation has to be solved which is closely related to Fibonacci numbers. The results obtained are extended to linear dif\/ferential equations of arbitrary order in Section~\ref{section5}. However, since the method used for the second order case turns out to be rather cumbersome in the general case (involving recurrence relations related to $n$-generalized Fibonacci numbers) we follow a~slightly dif\/ferent approach. The main dif\/f\/iculty consists in getting control over the case where degeneracies appear in the associated characteristic equation. In Section~\ref{section6} the general system of f\/irst order is considered. It is rather simple to describe the solution of this system abstractly but when considering concrete examples tedious calculations result, too. The peculiarities of the linear dif\/ferential equation with variable coef\/f\/icients are sketched in Section~\ref{section7} by considering an example. In Section~\ref{section8} a very simple nonlinear dif\/ferential equation is treated. It turns out that -- in contrast to the linear dif\/ferential equations where a close analogy to the ``bosonic'' case holds true -- due to the nilpotency of the para-Grassmann variable the equations show a behaviour reminding one also of ``bosonic'' partial dif\/ferential equations. Finally, some conclusions are presented in Section~\ref{section9}.

\section{Some formulae of para-Grassmann calculus}\label{section2}

In this section we brief\/ly recall some basic facts about para-Grassmann
calculus following \cite{matthias} (see also \cite{ABC,FIK2,FIK1,FIK3}). For a given natural number $p\geq 1$ the nilpotent
``coordinate'' $\theta$ vanishes in $(p+1)$-th order, i.e., $\theta^{p+1}=0$ (with $\theta^l \neq 0$ for $0\leq l \leq p$); the case $p=1$ reproduces
the well known Grassmann variable appearing in supersymmetric
theories. Due to our assumption there will be the linearly independent
(over $\mathbb{C}$) elements
$\{1,\theta,\theta^2,\ldots,\theta^p\}$. In analogy to ``ordinary'' supermathematics we introduce ``parasuperfunctions'' $f(x,\theta)$, which can be expanded as a f\/inite series in $\theta$:
\begin{gather}\label{exp}
f(x,\theta)=f_0(x)+f_1(x)\theta+\cdots+f_p(x)\theta^p.
\end{gather}
The functions $f_k(x)$ will be called the {\it components} (or sometimes {\it coefficients}) of $f(x,\theta)$. We introduce a derivative $\partial \equiv \partial_{\theta}$ satisfying the following commutation relation:
\begin{gather}\label{commurel}
\partial\theta=q\theta\partial,\qquad q=e^{\frac{2\pi i}{p+1}}.
\end{gather}
In the following we use the standard notation
\[[n]_q=1+q+\cdots+q^{n-1}\equiv \frac{1-q^n}{1-q},\qquad
[n]_q!=[n]_q[n-1]_q\cdots [1]_q
\]
with $[0]_q=[0]_q!=1$. It is easy to check that the commutation relation \eqref{commurel} implies $\partial
\theta=1$, $\partial \theta^2=(1+q)\theta$, and in general
\[
\partial\theta^n=[n]_q\theta^{n-1}.
\]
Note that $\partial^{p+1}=0$. For parasuperfunctions we
have (at least) two derivatives, the ``normal'' derivative
$\partial_x\equiv \frac{d}{dx}$ and $\partial$ acting on the nilpotent
variables. We introduce a parasupercovariant derivative by \cite{ABC,FIK2,FIK1,FIK3}
\begin{gather}\label{paracov}
{\mathcal{D}}:=\partial + \frac{\theta^p}{[p]_q!} \partial_x.
\end{gather}
It is straightforward to check that for $f$ as in \eqref{exp} one has (for $0\leq r \leq
p $)
\begin{gather}\label{covderi}
{\mathcal{D}}^{r} f(x,\theta)=\sum_{k=0}^{r-1}
\frac{[k]_q!}{[p-(r-1)+k]_q!}\partial_xf_k(x)\theta^{p-(r-1)+k} +\sum_{k=r}^p \frac{[k]_q!}{[k-r]_q!}f_k(x)\theta^{k-r}.
\end{gather}
Note in particular that the parasupercovariant derivative $\mathcal{D}$
is a $(p+1)$-th root of
$\partial_x$ in the sense that \cite{ABC,FIK2,FIK1,FIK3}
\begin{gather}\label{rootp}
{\mathcal{D}}^{p+1} f(x,\theta)=\partial_x f(x,\theta).
\end{gather}
Due to this relation there exists a certain ``periodicity'' in the
derivatives ${\mathcal{D}}^{s} f $ of a function $f$ where now $s$ is
an arbitrary integer. Since $s$ can be decomposed in a unique way as
$s=s'(p+1)+s''$ with $0\leq s'' <p+1$, one may use \eqref{covderi}
and \eqref{rootp} to obtain
\begin{gather}
{\mathcal{D}}^{s} f(x,\theta)=\sum_{k=0}^{s''-1}
\frac{[k]_q!}{[p-(s''-1)+k]_q!}\partial_x^{s'+1}f_k(x)\theta^{p-(s''-1)+k}\nonumber\\
\phantom{{\mathcal{D}}^{s} f(x,\theta)=}{} +\sum_{k=s''}^p
\frac{[k]_q!}{[k-s'']_q!}\partial_x^{s'} f_k(x)\theta^{k-s''}.\label{genabl}
\end{gather}
It is also possible to introduce a formal ``antiderivative'' by letting \cite{FRausch}
\[
{\mathcal{D}}^{-1}f(x,\theta):=[p]_q!\int^x f_p(t)\, dt +
f_0(x)\theta+\frac{1}{[2]_q}f_1(x)\theta^2+\cdots+ \frac{1}{[p]_q}f_{p-1}(x)\theta^p
\]
so that ${\mathcal{D}}^{-1}{\mathcal{D}}f=f $; in particular, ${\mathcal{D}}^{-1}1=\theta$. More generally,
iterating this allows one to introduce an operator ${\mathcal{D}}^{-s}
$. Let us also introduce an exponential function in a bosonic and a~para-Grassmann variable by setting \cite{FIK2}
\begin{gather}\label{expo}
e_q(x;\theta):=e^x \sum_{n=0}^p \frac{\theta^n}{[n]_q!}.
\end{gather}
Then it is an easy calculation to verify that it has the def\/ining
property of the exponential function, i.e.,
${\mathcal{D}}e_q(x;\theta)=e_q(x;\theta)$. Under a
scaling of the arguments we obtain the following nice expression \cite{FIK2}:
\[
{\mathcal{D}}e_q(\beta^{p+1} x;\beta
\theta)=\beta e_q(\beta^{p+1} x;\beta
\theta).
\]
This implies ${\mathcal{D}}^{p+1}e_q(\beta^{p+1} x;\beta
\theta)=\partial_x
e_q(\beta^{p+1}x;\beta \theta)$, as it should according to
\eqref{rootp}.

\section{Some very simple dif\/ferential equations}\label{section3}

In this brief section we consider some simple dif\/ferential equations. Most of these results can be found already in \cite{matthias} but for the convenience of the reader and the systematic exposition we have included the results here. Now, let us consider the very simple dif\/ferential equation
\begin{gather}\label{dgl1}
{\mathcal{D}}^sf(x,\theta)=0.
\end{gather}
Writing $s=s'(p+1)+s''$ with $0\leq s'' < p+1$, we  may use
\eqref{genabl} to obtain the following conditions on the components
$f_k(x)$:
\[
\partial_{x}^{s'+1}f_k(x)=0, 0\leq k \leq s''-1,\qquad \partial_{x}^{s'}f_k(x)=0, s''\leq k \leq p.
\]
Thus, the f\/irst $s''$ components are polynomials of degree $s'$,
whereas the remaining components are polynomials of degree
$s'-1$. In particular, there are
$s''(s'+1)+(p+1-s'')s'=s'(p+1)+s''\equiv s$ free parameters for a
solution $f$ of \eqref{dgl1}, showing that
the kernel of ${\mathcal{D}}^s$ has dimension $s$. In the case $s=1$ this means that the only solution of ${\mathcal{D}}f=0$ is given by $f(x,\theta)=c$ for some $c\in \mathbb{C}$. Thus, the ``parasupercovariant constant function'' (meaning that ${\mathcal{D}}f=0$) is indeed given by a~constant value $f(x,\theta)=c$. In the case $s=2$ the ``Laplace equation'' ${\mathcal{D}}^2f=0$ in one dimension has two linearly
independent solutions. Consider the inhomogeneous version of \eqref{dgl1}, i.e.,
\begin{gather}\label{dgl1a}
{\mathcal{D}}^sf(x,\theta)=g(x,\theta).
\end{gather}
A particular solution is given by ${\mathcal{D}}^{-s}g$. For any solution $f_{\rm hom}$ of the homogeneous equation the sum $f_{\rm hom}+{\mathcal{D}}^{-s}g$ is a solution of the inhomogeneous equation. Thus, the set of solutions is
an af\/f\/ine space (as in the usual ``bosonic'' case). Let us summarize these observations in the following proposition.

\begin{proposition}\label{proposition1} Let the differential equation \eqref{dgl1} with $s=s'(p+1)+s''$ where $0\leq s'' < p+1$ be given. Any solution of \eqref{dgl1} can be written as
\[
f_{\rm hom}(x,\theta)=\sum_{k=0}^{s''-1}\left\{\sum_{r=0}^{s'}c_{k,r}x^r\right\}\theta^k
+\sum_{k=s''}^{p}\left\{\sum_{r=0}^{s'-1}c_{k,r}x^r\right\}\theta^k,
\]
where $c_{k,r}\in \mathbb{C}$. Thus, the space of solutions is a complex linear space of dimension $s$. The associated inhomogeneous differential equation~\eqref{dgl1a} has the particular solution ${\mathcal{D}}^{-s}g$. For any solution $f_{hom}$ of the homogeneous equation the sum $f_{\rm hom}+{\mathcal{D}}^{-s}g$ is a solution of the inhomogeneous equation. Thus, the set of solutions of \eqref{dgl1a} is
a complex affine space of dimension~$s$.
\end{proposition}

Now, let us turn to the ``eigenvalue problem''
\begin{gather}\label{DE}
{\mathcal{D}}f(x,\theta)=\lambda f(x,\theta)
\end{gather}
for $\lambda \in \mathbb{C}$. From above we know that $f(x,\theta)=Ce_q(\lambda^{p+1}x;\lambda
\theta)$ is a solution. To f\/ind out whether there exists another
solution, we write $f(x,\theta)=\sum_{k=0}^pf_k(x)\theta^k$. Inserting this into \eqref{DE} gives for $k=0,\ldots,
p-1 $ the recursion relation
$f_{k+1}(x)=\frac{\lambda}{[k+1]_q}f_k(x)$ as well as $\partial_x
f_0(x)=\lambda [p]_q!f_p(x)$. Using the recursion relation repeatedly yields
$f_p=\frac{\lambda^p}{[p]_q!}f_0$. Using now the second equation yields
the dif\/ferential equation $\partial_x f_0(x)=\lambda^{p+1}f_0(x)$ for
$f_0$ with the solution $f_0(x)=Ce^{\lambda^{p+1}x}$ where $C\in \mathbb{C}$ is a constant. From the recursion
relation it follows that
$f_k(x)=\frac{\lambda^k}{[k]_q!}Ce^{\lambda^{p+1}x}$ and, consequently, that $f(x,\theta)= Ce_q(\lambda^{p+1}x;\lambda \theta)$
is indeed the general solution of the dif\/ferential equation \eqref{DE}. Using
\eqref{genabl}, it is straightforward to consider the slightly more dif\/f\/icult eigenvalue problem
\begin{gather}\label{easy}
{\mathcal{D}}^sf(x,\theta)=\lambda
f(x,\theta), \qquad s \in {\bf N}.
\end{gather}
Namely, let $\lambda^{\frac{1}{s}}$ be one of the $s$
roots satisfying $(\lambda^{\frac{1}{s}})^s=\lambda$. All $s$ roots
are then given by
\[\big\{\lambda^{\frac{1}{s}},
\mu\lambda^{\frac{1}{s}},\ldots,\mu^{s-1}\lambda^{\frac{1}{s}}\big\},
\]
where $\{1,\mu,\mu^2,\ldots,\mu^{s-1}\}$ is the cyclic group of order
$s$ (isomorphic to ${\bf Z}_s$) consisting of the $s$ roots of
unity. Since each function $e_q((\mu^k\lambda^{\frac{1}{s}})^{p+1}x;(\mu^k\lambda^{\frac{1}{s}})\theta)$
solves \eqref{easy}, the solution of \eqref{easy} is given by $f(x,\theta)=\sum_{k=0}^{s-1}C_k
e_q((\mu^k\lambda^{\frac{1}{s}})^{p+1}x;(\mu^k\lambda^{\frac{1}{s}})\theta)$. The space of solutions has dimension
$s$, as expected. Let us summarize these observations in the following proposition.

\begin{proposition} \label{proposition2} Let the differential equation \eqref{easy} be given. Let $\lambda^{\frac{1}{s}}$ be one of the $s$
roots satisfying $(\lambda^{\frac{1}{s}})^s=\lambda$ and let $\{1,\mu,\mu^2,\ldots,\mu^{s-1}\}$ be the cyclic group of order~$s$. Any solution of~\eqref{easy} can be written as
\[
f(x,\theta)=\sum_{k=0}^{s-1}C_k
e_q\big(\big(\mu^k\lambda^{\frac{1}{s}}\big)^{p+1}x;\big(\mu^k\lambda^{\frac{1}{s}}\big)\theta\big),
\]
where $C_k\in \mathbb{C}$. In particular, the set of solutions of \eqref{easy} is a complex linear space of dimension~$s$.
\end{proposition}

Up to now, we have only considered equations where a
simple ansatz using the exponential function \eqref{expo} yielded all solutions. In the following
sections more complicated expressions in ${\mathcal{D}} $ will be
discussed. Due to the lack of a general product rule these cases will
turn out to be more dif\/f\/icult and it is unclear what a ``good'' ansatz
should be. This will appear already in the case of linear dif\/ferential equations of second order discussed in the next section.

\begin{remark} \label{remark1} Let us denote the set of parasuperfunctions $f\equiv f(x,\theta)$ by $\mathcal{A}$ (where we assume for ease of presentation that the components $f_k$ are ``suf\/f\/iciently nice'', e.g., in $C^{\infty}$). Clearly, it is a linear space (over $\mathbb{C}$) and if we also consider as product of elements $f,g\in \mathcal{A}$ the pointwise product (i.e., $(fg)(x,\theta):=f(x,\theta)g(x,\theta)$) it is even a commutive ring with unit (having nilpotent elements). The parasupercovariant ``derivative'' $\mathcal{D}:\mathcal{A}\rightarrow \mathcal{A}$ is, however, not a derivation in this ring, i.e., $\mathcal{D}(fg)\neq \mathcal{D}(f)g+f\mathcal{D}(g)$. Thus, the structure $(\mathcal{A},\cdot,\mathcal{D})$ is not a dif\/ferential ring as considered, e.g., in \cite{put} and the literature given therein. It is tempting to introduce a~new product~$*$ such that $(\mathcal{A},*,\mathcal{D})$ becomes a dif\/ferential ring, i.e., \mbox{$\mathcal{D}(f*g)= \mathcal{D}(f)*g+f*\mathcal{D}(g)$}. However, these conditions on the new product seem to be rather involved and a natural interpretation is lacking.
\end{remark}

\section{The linear dif\/ferential equation of second order}\label{section4}

In this section we will discuss the general linear dif\/ferential
equation of second order. More precisely, we will show the following theorem:
\begin{theorem}\label{second}
Let $c_1,c_2\in \mathbb{C}$; the general solution of
\begin{gather}\label{2deg}
({\mathcal{D}}^2+c_1{\mathcal{D}}+c_2)f(x,\theta)=0
\end{gather}
may be obtained as follows. Define $\lambda_{\pm}:=-\frac{c_1}{2}\pm
\frac{1}{2}\sqrt{c_1^2-4c_2}$. In the non-degenerated case where
$\lambda_{+} \neq \lambda_{-}$ the general solution of \eqref{2deg} is given by
\[f(x,\theta)=C_1e_q(\lambda_{+}^{p+1}x;\lambda_{+}\theta)+C_2e_q(\lambda_{-}^{p+1}x;\lambda_{-}\theta)
\]
with arbitrary $C_i\in \mathbb{C}$. In the degenerated case where $c_1^2=4c_2$ we abbreviate $\alpha:=-\frac{c_1}{2}$ $($thus $c_2=\alpha^2)$; the general solution is in this case given by
\begin{gather}\label{eqsecond}
f(x,\theta)=C_1e_q(\alpha^{p+1}x;\alpha\theta)+
C_2\left((p+1)\alpha^pxe_q(\alpha^{p+1}x;\alpha\theta)
+e^{\alpha^{p+1}x}\sum_{k=1}^p\frac{k\alpha^{k-1}\theta^k}{[k]_q!} \right).
\end{gather}
\end{theorem}

Note that making an ansatz of exponential form (as we did in the last
section) yields in the non-degenerated case both solutions and in the
degenerated case the f\/irst solution. Before giving a proof of the
theorem, we single out the cases of smallest $p$ explicitly.
\begin{corollary} \label{corollary1} Consider the degenerated case $({\mathcal{D}}^2
-2\alpha{\mathcal{D}}+\alpha^2)f(x,\theta)=0 $. A first solution is
always $($i.e., for arbitrary $p)$ given by $f_1(x,\theta)=e_q(\alpha^{p+1}x;\alpha\theta)$. A linearly
independent solution is given in the case $p=1$ $($i.e., for an
``ordinary'' Grassmann variable$)$ by
\[
f_2(x,\theta)=2\alpha x e^{\alpha^2x}+\big(1+2\alpha^2x\big)e^{\alpha^2x}\theta.
\]
In the case $p=2$ a second solution is given by
\[
f_2(x,\theta)=3\alpha^2xe^{\alpha^3x}+\big(1+3\alpha^3x\big)e^{\alpha^3x}\theta+\frac{\alpha}{[2]_q!}
\big(2+3\alpha^3x\big)e^{\alpha^3x}\theta^2.
\]
\end{corollary}
Let us now turn to the proof of the theorem. Instead of just inserting the claimed solutions and see whether they fulf\/ill the dif\/ferential equation we will give a longer proof and show how these solutions are found.

\begin{proof}[Prof of Theorem \ref{second}]
Let $f(x,\theta)=f_0(x)+f_1(x)\theta+\cdots +
f_p(x)\theta^p$ be a solution of (\ref{2deg}). Inserting this and
comparing coef\/f\/icients of equal power in $\theta$ yields the
following system of equations:
\begin{gather}\label{rec}
f_{k+2}(x) = -c_1\frac{1}{[k+2]_q}f_{k+1}(x)-c_2\frac{1}{[k+2]_q[k+1]_q}f_k, \qquad 0 \leq k \leq p-2,\\ %\label{recdeg}
\partial_xf_0(x) = -c_1[p]_q!f_p(x)-c_2[p-1]_q!f_{p-1}(x),\label{recdeg2} \\
\partial_xf_1(x) = -c_1\partial_xf_0(x) -c_2[p]_q!f_{p}(x).\label{recdeg2f1}
\end{gather}
Note that in the case $p=1$ the recursion relations \eqref{rec} are vacuous. Only the dif\/ferential equations \eqref{recdeg2}--\eqref{recdeg2f1} remain and can be cast into the form
\begin{gather}\label{diffp1}
\left(\begin{array}{c} \partial_xf_0(x)\\ \partial_xf_{1}(x)\end{array}\right)=\left(\begin{array}{ll}
-c_2 &-c_1 \\ c_1c_2 &(c_1^2-c_2)\end{array}\right)\left(\begin{array}{c} f_0(x)\\ f_{1}(x)\end{array}\right).
\end{gather}
Let us return to the case of arbitrary $p$. Here we will f\/irst solve the recursion relations until we arrive at a similar system of dif\/ferential equations for the components $f_0(x)$ and $f_1(x)$ which can be solved. To solve the recursion relations, we introduce for $k=0,\ldots,p-1$ the vectors and matrices
\begin{gather}\label{notat}
v_k(x):= \left(\begin{array}{c} f_k(x)\\ f_{k+1}(x)\end{array}
\right),\qquad \TT_{k+1}:= \left(\begin{array}{cc} 0 & 1\\ -
\frac{c_2}{[k+2]_q[k+1]_q}& -\frac{c_1}{[k+2]_q}\end{array}
\right).
\end{gather}
This enables us to write the recursion relation \eqref{rec} in the compact
form $v_{k+1}(x)=\TT_{k+1}v_{k}(x)$. Iterating this yields
\begin{gather}\label{iter}
v_{k+1}(x)=\TT_{k+1}\TT_{k}\cdots \TT_1 v_{0}(x)= \FF_{k+1}v_0(x).
\end{gather}
Since $v_{p-1}(x)= \FF_{p-1}v_0(x)$, this allows us to express $f_{p-1}$ and $f_p$ through $f_0$ and $f_1$ once we have determined $\FF_{p-1}$. Inserting this in \eqref{recdeg2} and \eqref{recdeg2f1} will give a system of ordinary
dif\/ferential equations for $f_0$, $f_1$. We will now determine $\FF_k$
for all $k$ since after having determined $f_0$ and $f_1$ the remaining components
$f_k$ are calculated using \eqref{iter}. Let us write the matrix
$\FF_k$ explicitly~as
\begin{gather}\label{2mat}
\FF_{k}= \left(\begin{array}{cc} \frac{f^{(1)}_{k-1}}{[k]_q!} &\frac{f^{(2)}_{k-1}}{[k]_q!} \vspace{1mm}\\ \frac{f^{(1)}_{k}}{[k+1]_q!} &\frac{f^{(2)}_{k}}{[k+1]_q!} \end{array}
\right).
\end{gather}
Note that we can then write
\begin{gather}\label{comp}
f_{k}(x)=\frac{f^{(1)}_{k-1}}{[k]_q!}f_0(x)+\frac{f^{(2)}_{k-1}}{[k]_q!}f_1(x).
\end{gather}
The relation $\FF_{k+1}=\TT_{k+1}\FF_{k}$ is equivalent to
\begin{gather}\label{fibo}
f^{(i)}_{k+2}=-c_1f^{(i)}_{k+1}-c_2f^{(i)}_{k}, \qquad i=1,2.
\end{gather}
In general, a solution of a generalized Fibonacci sequence
$f_{k+2}=-c_1f_{k+1}-c_2f_k$ is obtained as follows: The generating
function $x^2+c_1x+c_2$ has zeroes $\lambda_{\pm}=-\frac{c_1}{2}\pm
\frac{1}{2}\sqrt{c_1^2-4c_2}$, so that by Binet's formula the general
term is given by $f_k=a_1 \lambda_{+}^k+a_2 \lambda_{-}^k$, where the
coef\/f\/icients $a_1$, $a_2$ are determined using the (given) initial values $f_0$, $f_1$. Note that we have the following relations for the coef\/f\/icients $c_i$:
\begin{gather}\label{coeff}
c_1 = -(\lambda_+ + \lambda_-), \qquad c_2=\lambda_+\lambda_-.
\end{gather}

{\it Let us now assume that $\lambda_{+}=\lambda_{-}$, i.e.,
$c_1^2=4c_2$}. We def\/ine $c_1=-2\alpha$, so that $c_2=\alpha^2$. In
this case we obtain (for $k=0,\ldots,p-1$)
\begin{gather}\label{degcase}
\FF_k=\left(\begin{array}{cc} -\frac{(k-1)\alpha^k}{[k]_q!}&\frac{k\alpha^{k-1}}{[k]_q!} \vspace{1mm}\\ -\frac{k\alpha^{k+1}}{[k+1]_q!}&\frac{(k+1)\alpha^k}{[k+1]_q!} \end{array}
\right).
\end{gather}
Recalling \eqref{iter}, we may now use this for $k=p-1$ to express $f_p$, $f_{p-1}$ through $f_0,f_1$. Inserting this into \eqref{recdeg2} and \eqref{recdeg2f1} gives the dif\/ferential equation $v_0'(x)=Av_0(x)$, where the matrix $A$ is
given by
\[
A= \left(\begin{array}{cc} -p\alpha^{p+1} & (p+1)\alpha^p \\
-(p+1)\alpha^{p+2}&(p+2)\alpha^{p+1}\end{array}\right).
\]
A has only one eigenvalue $\lambda=\alpha^{p+1}$ with eigenvector $(1\,\,\, \alpha)^t$. Changing to the appropriate basis allows us to write
\[
\frac{d}{dx} \left(\begin{array}{c} f_0(x) \\ f_1(x)+\alpha
f_0(x)\end{array}\right) = \left(\begin{array}{cc} \alpha^{p+1} &
(p+1)\alpha^p \\ 0 & \alpha^{p+1}\end{array}\right)
\left(\begin{array}{c}f_0(x)\\ f_1(x)+\alpha
f_0(x)\end{array}\right),
\]
which implies that there exist $C_1,C_2\in\mathbb{C}$ such that
\begin{gather*}
f_0(x) =(C_1+x(p+1)\alpha^p C_2)e^{\alpha^{p+1}x},\\
f_1(x) =(\alpha C_1+x(p+1)\alpha^{p+1}C_2+C_2)e^{\alpha^{p+1}x}.
\end{gather*}
On the other hand, by \eqref{degcase} we obtain that
\[
f_k(x)=\frac{k\alpha^k}{[k]_q!}(\frac{1}{\alpha}f_1(x)-f_0(x))+\frac{\alpha^k}{[k]_q!}f_0(x)
\]
for $k=2,3,\ldots,p$. Hence,
\begin{gather*}
f(x,\theta) =f_0(x)+f_1(x)\theta+\sum_{k=2}^pf_k(x)\theta^k
 =f_0(x)\sum_{k=0}^p\frac{(\alpha\theta)^k}{[k]_q!}+\frac{f_1(x)-\alpha f_0(x)}{\alpha}\sum_{k=0}^p\frac{k(\alpha\theta)^k}{[k]_q!}\\
\phantom{f(x,\theta)}{} =C_1e^{\alpha^{p+1}x}\sum_{k=0}^p\frac{(\alpha\theta)^k}{[k]_q!}+C_2e^{\alpha^{p+1}x}\left(
\sum_{k=1}^p\frac{k\alpha^{k-1}\theta^k}{[k]_q!}+x(p+1)\alpha^p\sum_{k=0}^p\frac{(\alpha\theta)^k}{[k]_q!}\right)\\
\phantom{f(x,\theta)}{}=C_1e_q(\alpha^{p+1}x;\alpha\theta)+
C_2\left((p+1)\alpha^pxe_q(\alpha^{p+1}x;\alpha\theta)+e^{\alpha^{p+1}x}\sum_{k=1}^p\frac{k\alpha^{k-1}\theta^k}{[k]_q!} \right),
\end{gather*}
which completes the proof of this case.

{\it We now assume that $\lambda_{+}\neq \lambda_{-}$}. Thus, we obtain the two generalized Fibonacci
sequences $(f^{(1)}_k)_{k\in \mathbb{N}}$ and $(f^{(2)}_k)_{k\in \mathbb{N}}$ which satisfy the
same recursion relation (\ref{fibo}) but with dif\/ferent initial
conditions. More precisely, one has for the f\/irst sequence
$f^{(1)}_0=0$ and $f^{(1)}_1=-c_2$, while the initial values for the
second sequence are given by $f^{(2)}_0=1$ and
$f^{(2)}_1=-c_1$. Def\/ining $\gamma := \sqrt{c_1^2-4c_2}$, we can write the general terms of these sequences with the help of Binet's formula as
\begin{gather}
f_k^{(1)}=\frac{c_2}{\gamma}\mu_k\equiv\frac{c_2}{\gamma}\big(\lambda_{-}^k-\lambda_{+}^k\big),
\nonumber\\
 f_k^{(2)}=\frac{1}{2}\nu_k+\frac{c_1}{2\gamma}\mu_k\equiv\frac{1}{2}
\left(1-\frac{c_1}{\gamma}\right)\lambda_{+}^k+\frac{1}{2}\left(1+\frac{c_1}{\gamma}\right)\lambda_{-}^k,\label{eqkkff}
\end{gather}
thereby def\/ining implicitly $\mu_k$, $\nu_k$. In particular, using this for $k=p-2$ in \eqref{iter} yields
\begin{gather}\label{eqffp}
f_{p-1}(x)=\frac{f^{(1)}_{p-2}}{[p-1]_q!}f_0(x)+\frac{f^{(2)}_{p-2}}{[p-1]_q!}f_1(x),\qquad
f_{p}(x)=\frac{f^{(1)}_{p-1}}{[p]_q!}f_0(x)+\frac{f^{(2)}_{p-1}}{[p]_q!}f_1(x).
\end{gather}
Combining \eqref{recdeg2}, \eqref{recdeg2f1} and \eqref{eqffp} gives the dif\/ferential equation
\[
v_0'(x)=\left(\begin{array}{c} \partial_xf_0(x)\\ \partial_xf_{1}(x)\end{array}\right)=Bv_0(x),
\]
where the matrix $B$ is given by
\begin{align*}
B&=\left(\begin{array}{ll}
-c_1f_{p-1}^{(1)}-c_2f_{p-2}^{(1)}&-c_1f_{p-1}^{(2)}-c_2f_{p-2}^{(2)}\\
(c_1^2-c_2)f_{p-1}^{(1)}+c_1c_2f_{p-2}^{(1)}&(c_1^2-c_2)f_{p-1}^{(2)}+c_1c_2f_{p-2}^{(2)}
\end{array}\right).
\end{align*}
Note that this is the generalization of \eqref{diffp1} for arbitrary $p$. Using \eqref{fibo} as well as \eqref{eqkkff}, we obtain that
\[
B=\left(\begin{array}{ll}
\frac{c_2}{\gamma}\mu_p&\frac{1}{2}\nu_p+\frac{c_1}{2\gamma}\mu_p\\
\frac{c_2}{\gamma}\mu_{p+1}&\frac{1}{2}\nu_{p+1}+\frac{c_1}{2\gamma}\mu_{p+1}
\end{array}\right)=\frac{1}{\gamma}\left(\begin{array}{ll}
c_2\mu_p&-\mu_{p+1}\\
c_2\mu_{p+1}&-\mu_{p+2}
\end{array}\right).
\]
In order to f\/ind the eigenvalues of the matrix $B$ let us f\/irst f\/ind the characteristic polynomial of~$B$:
\begin{gather*}
\det(B-xI) = x^2-\frac{1}{\gamma}(c_2\mu_p-\mu_{p+2})x+\frac{c_2}{\gamma^2}\big(\mu_{p+1}^2-\mu_{p}\mu_{p+2}\big).
\end{gather*}
Recalling \eqref{coeff} and $\gamma=\lambda_+-\lambda_-$, we obtain
that
\begin{eqnarray*}
\det(B-xI)&=&x^2-\nu_{p+1}x+c_2^{p+1}=\big(x-\lambda_{+}^{p+1}\big)\big(x-\lambda_{-}^{p+1}\big).
\end{eqnarray*}
Hence, the eigenvalues of the matrix $B$ are $\lambda_{-}^{p+1}$ and
$\lambda_{+}^{p+1}$ with the eigenvectors $(1\,\,  \lambda_{-})^t$ and $(1 \,\, \lambda_+)^t$, respectively. Therefore, there exist constants $C_i\in\mathbb{C}$, $i=1,2$,
such that
\begin{gather}\label{eqf01}
\left(\begin{array}{l}f_0(x)\\f_1(x)\end{array}\right)
=C_1e^{\lambda_{-}^{p+1}x}\left(\begin{array}{l}1\\
\lambda_{-}\end{array}\right)+C_2e^{\lambda_{+}^{p+1}x}\left(\begin{array}{l}1\\
\lambda_{+}\end{array}\right).
\end{gather}
In the case $p=1$ we are f\/inished since all components $f_0(x)$ and $f_1(x)$ have been determined. For arbitrary $p$ we use \eqref{comp} and \eqref{eqkkff} to get that
\begin{gather*}
f_k(x)  =  \frac{\frac{c_2}{\gamma}\mu_{k-1}}{[k]_q!}f_0(x)
+\frac{\frac{1}{2}\nu_{k-1}+\frac{c_1}{2\gamma}\mu_{k-1}}{[k]_q!}f_1(x)
 = \frac{1}{\gamma[k]_q!}(c_2\mu_{k-1}f_0(x)-\mu_kf_1(x))
\end{gather*}
for all $k=2,3,\ldots,p$. This in turn yields
\begin{gather*}
f(x,\theta) =\sum_{k=0}^pf_k(x)\theta^k
 =f_0(x)+f_1(x)\theta+\sum_{k=2}^p\frac{\theta^k}{\gamma[k]_q!}(c_2\mu_{k-1}f_0(x)-\mu_kf_1(x))\\
\phantom{f(x,\theta)}{}
=f_0(x)+f_1(x)\theta+\frac{c_2f_0(x)}{\gamma}\sum_{k=2}^p\frac{\mu_{k-1}\theta^k}{[k]_q!}
-\frac{f_1(x)}{\gamma}\sum_{k=2}^p\frac{\mu_k\theta^k}{[k]_q!}\\
\phantom{f(x,\theta)}{}
=f_0(x)+f_1(x)\theta+\frac{\lambda_{+}f_0(x)-f_1(x)}{\gamma}\sum_{k=2}^p\frac{(\lambda_{-}\theta)^k}{[k]_q!}
+\frac{f_1(x)-\lambda_{-}f_0(x)}{\gamma}\sum_{k=2}^p\frac{(\lambda_{+}\theta)^k}{[k]_q!}.
\end{gather*}
Using \eqref{eqf01}, we f\/ind that
\[
f_0(x)+f_1(x)\theta = C_1e^{\lambda_{-}^{p+1}x}(1+\lambda_{-}\theta)+C_2e^{\lambda_{+}^{p+1}x}(1+\lambda_{+}\theta)
\]
as well as $\lambda_{+}f_0(x)-f_1(x)=\gamma C_1e^{\lambda_{-}^{p+1}x}$ and
$f_1(x)-\lambda_{-}f_0(x)=\gamma C_2e^{\lambda_{+}^{p+1}x}$, implying
\begin{gather*}
f(x,\theta)
=C_1e^{\lambda_{-}^{p+1}x}(1+\lambda_{-}\theta)+C_2e^{\lambda_{+}^{p+1}x}
(1+\lambda_{+}\theta)\\
\phantom{f(x,\theta)=}{} +C_1e^{\lambda_{-}^{p+1}x}
\sum\limits_{k=2}^p\frac{(\lambda_{-}\theta)^k}{[k]_q!}+C_2e^{\lambda_{+}^{p+1}x}
\sum\limits_{k=2}^p\frac{(\lambda_{+}\theta)^k}{[k]_q!}\\
\phantom{f(x,\theta)}{}=C_1e^{\lambda_{-}^{p+1}x}\sum\limits_{k=0}^p\frac{(\lambda_{-}\theta)^k}{[k]_q!}
+C_2e^{\lambda_{+}^{p+1}x}\sum\limits_{k=0}^p\frac{(\lambda_{+}\theta)^k}{[k]_q!}\\
\phantom{f(x,\theta)}{}
=C_1e_q(\lambda_{-}^{p+1}x,\lambda_{-}\theta)+C_2e_q(\lambda_{+}^{p+1}x,\lambda_{+}\theta),
\end{gather*}
which completes the proof.
\end{proof}

\section{The linear dif\/ferential equation of arbitrary order}\label{section5}

In this section we solve the linear dif\/ferential equation of order $n$, where we assume for ease of presentation that $n\leq p$ (for $n \geq p+1$ one has to consider \eqref{genabl} instead
of \eqref{covderi}). Thus, we are considering the equation
\begin{gather}\label{deggen}
({\mathcal{D}}^n+c_1{\mathcal{D}}^{n-1}+ \cdots + c_{n-1}{\mathcal{D}}+c_n)f(x,\theta)=0.
\end{gather}
Observe that the restriction $n \leq p$ means that for $p=1$ we only consider $n\leq 1$ -- but this case was already considered in the above sections. Thus, the case $p=1$ doesn't lead to new cases here. At the end of the section we consider brief\/ly what happens when $n>p$ (see Proposition~\ref{propgen}).

The strategy for the solution of the above dif\/ferential equation consists of two steps: In the f\/irst step we consider the linear dif\/ferential equation associated to one degenerated eigenvalue and in the second step we assemble these results for the dif\/ferential equation \eqref{deggen}. The second step is only of a formal nature and the main work consists in f\/inding the solutions of the degenerated dif\/ferential equation.

In the following we will need multi-indices. Therefore, let us f\/ix the notation. We write
\[
{\bf m}_s :=(m_0,m_1,\dots,m_s)
\]
where $m_0\geq 0$ and $m_i\geq 1$ for $i=1,2,\ldots,s$ are natural numbers. We also introduce the length of ${\bf m}_s$ by
\[
|{\bf m}_s|:=\sum_{i=0}^s m_i.
\]

\begin{lemma}\label{lemmacompo}
Let $f(x,\theta)=\sum_{k=0}^pf_k(x)\theta^k$ be a solution of the differential equation
\[
({\mathcal{D}}-\lambda)^nf(x,\theta)=0,\qquad 1 \leq n\leq p .
\]
Then there exist $C_1,\ldots,C_n\in\mathbb{C}$ such that
\begin{gather}\label{compo}
f_k(x)=e^{\lambda^{p+1}x}\frac{\lambda^k}{[k]_q!}\sum_{s=0}^{n-1}\left\{\sum_{|{\bf m}_s|\leq n-1}C_{n-|{\bf m}_s|}\lambda^{s-|{\bf m}_s|}\binom{k}{m_0}\prod_{i=1}^s\binom{p+1}{m_i}\right\}\frac{(\lambda^p x)^s}{s!},
\end{gather}
where ${\bf m}_s=(m_0,m_1,\dots,m_s)$ with $m_0\geq0$ and $m_i\geq1$ for all $i=1,2,\ldots,s$.
\end{lemma}
\begin{proof}
We proceed the proof by induction on $n\geq1$. For $n=1$, the lemma gives -- due to $s=0$ and, hence, $m_0=0$ -- that
$f_k(x)=C_1e^{\lambda^{p+1}x}\frac{\lambda^k}{[k]_q!}$ which agrees with the solution of \eqref{DE}. Thus, the lemma holds for $n=1$. Now, assume that the lemma holds for $n$, and let us prove it for $n+1$. Let us write
\[
({\mathcal{D}}-\lambda)^{n+1}f(x,\theta)=({\mathcal{D}}-\lambda)^{n}g(x,\theta)=0
\]
with $g(x,\theta)=({\mathcal{D}}-\lambda)f(x,\theta)$. Using the induction hypothesis (for $g(x,\theta)$), we obtain that there exist
$C_1,\ldots,C_n\in\mathbb{C}$ such that
\[
\sum_{k=0}^pg_k(x)\theta^k=e^{\lambda^{p+1}x}\sum_{k=0}^p\frac{(\lambda\theta)^k}{[k]_q!}\sum_{s=0}^{n-1}\left\{\sum_{|{\bf m}_s|\leq n-1}C_{n-|{\bf m}_s|}\lambda^{s-|{\bf m}_s|}\binom{k}{m_0}\prod_{i=1}^s\binom{p+1}{m_i}\right\}\frac{(\lambda^p x)^s}{s!}.
\]
On the other hand, recalling $g(x,\theta)=({\mathcal{D}}-\lambda)f(x,\theta)$ and using \eqref{covderi}, we obtain
\[
\sum_{k=0}^pg_k(x)\theta^k=\frac{1}{[p]_q!}\partial_xf_0(x)\theta^p+\sum_{k=1}^p\frac{[k]_q!}{[k-1]_q!}f_k(x)\theta^{k-1}
-\lambda\sum_{k=0}^pf_k(x)\theta^k.
\]
By comparing the coef\/f\/icients of $\theta^k$ of the above two equations, we obtain that
\begin{gather}
g_p(x)=\frac{1}{[p]_q!}\partial_x f_0(x)-\lambda f_p(x),\nonumber\\
 g_k(x)=[k+1]_q f_{k+1}(x)-\lambda f_k(x),\qquad  1\leq k \leq p-1.\label{eqsf0}
\end{gather}
Therefore,
\begin{gather*}
\partial_xf_0(x)-\lambda^{p+1}f_0(x) =\partial_xf_0(x)+\lambda\sum_{i=0}^{p-1}\lambda^i[p-i]_q!f_{p-i}(x)
-\lambda\sum_{i=0}^p\lambda^i[p-i]_q!f_{p-i}(x)\\
\phantom{\partial_xf_0(x)-\lambda^{p+1}f_0(x)}{} =\sum_{i=0}^p\lambda^i[p-i]_q!g_{p-i}(x).
\end{gather*}
Noting $\partial_x(e^{-\lambda^{p+1}x}f_0(x))=e^{-\lambda^{p+1}x}(\partial_xf_0(x)-\lambda^{p+1}f_0(x))$, we conclude that
\[
\partial_x\big(e^{-\lambda^{p+1}x}f_0(x)\big)=e^{-\lambda^{p+1}x}\sum_{i=0}^p\lambda^i[p-i]_q!g_{p-i}(x).
\]
Inserting the expression for $g_{p-i}(x)$ given above, this implies
\begin{gather*}
e^{\lambda^{p+1}x}\partial_x(e^{-\lambda^{p+1}x}f_0(x))\\
=\sum_{i=0}^p\lambda^i[p-i]_q! e^{\lambda^{p+1}x}\frac{\lambda^{p-i}}{[p-i]_q!}\sum_{s=0}^{n-1}\left\{\sum_{|{\bf m}_s|\leq n-1}C_{n-|{\bf m}_s|}\lambda^{s-|{\bf m}_s|}\binom{p-i}{m_0}\prod_{j=1}^s\binom{p+1}{m_j}\right\}\frac{(\lambda^p x)^s}{s!}\\
=e^{\lambda^{p+1}x}\lambda^p\sum_{i=0}^p\sum_{s=0}^{n-1}\left\{\sum_{|{\bf m}_s|\leq n-1}C_{n-|{\bf m}_s|}\lambda^{s-|{\bf m}_s|}\binom{p-i}{m_0}\prod_{j=1}^s\binom{p+1}{m_j}\right\}\frac{(\lambda^p x)^s}{s!}.
\end{gather*}
Thus,
\[
\partial_x(e^{-\lambda^{p+1}x}f_0(x))=\lambda^p\sum_{i=0}^p\sum_{s=0}^{n-1}\left\{\sum_{|{\bf m}_s|\leq n-1}C_{n-|{\bf m}_s|}\lambda^{s-|{\bf m}_s|}\binom{p-i}{m_0}\prod_{j=1}^s\binom{p+1}{m_j}\right\}\frac{(\lambda^p x)^s}{s!}.
\]
Using
\[\frac{(\lambda^p x)^s}{s!}=\frac{1}{\lambda^p}\partial_x\left(\frac{(\lambda^p x)^{s+1}}{(s+1)!} \right),
\]
we have found
\[
\partial_x(e^{-\lambda^{p+1}x}f_0(x))=\partial_x\sum_{i=0}^p\sum_{s=0}^{n-1}\left\{\sum_{|{\bf m}_s|\leq n-1}C_{n-|{\bf m}_s|}\lambda^{s-|{\bf m}_s|}\binom{p-i}{m_0}\prod_{j=1}^s\binom{p+1}{m_j}\right\}\frac{(\lambda^p x)^{s+1}}{(s+1)!},
\]
or,
\[
e^{-\lambda^{p+1}x}f_0(x)=b+ \sum_{s=0}^{n-1}\left\{\sum_{i=0}^p\sum_{|{\bf m}_s|\leq n-1}C_{n-|{\bf m}_s|}\lambda^{s-|{\bf m}_s|}\binom{p-i}{m_0}\prod_{j=1}^s\binom{p+1}{m_j}\right\}\frac{(\lambda^p x)^{s+1}}{(s+1)!}
\]
for some constant $b \in \mathbb{C}$. Thus, using $\sum_{i=0}^{p-m_0}\binom{p-i}{m_0}=\binom{p+1}{m_0+1}$, we f\/ind
\begin{gather*}
e^{-\lambda^{p+1}x}f_0(x)=b+ \sum_{s=1}^{n}\!\left\{\sum_{|{\bf m}_{s-1}|\leq n-1}\!\!C_{n-|{\bf m}_{s-1}|}\lambda^{s-(|{\bf m}_{s-1}|+1)}\binom{p+1}{m_0+1}\prod_{j=1}^{s-1}\binom{p+1}{m_j}\!\right\}\frac{(\lambda^p x)^{s}}{s!}.
\end{gather*}
Introducing a new multi-index ${\bf n}_{s}=(n_0,n_1,\ldots,n_{s})$ by setting
\[
n_0:=0,\qquad n_1:=m_0+1, \qquad n_i:=m_{i-1},\qquad 2\leq i \leq s,
\]
we f\/ind $|{\bf m}_{s-1}|=|{\bf n}_{s}|-1$ and, consequently,
\[
e^{-\lambda^{p+1}x}f_0(x)=b+ \sum_{s=1}^{n}\left\{\sum_{|{\bf n}_{s}|\leq n; n_0=0}C_{n-|{\bf n}_{s}|+1}\lambda^{s-|{\bf n}_{s}|}\prod_{j=1}^{s}\binom{p+1}{n_j}\right\}\frac{(\lambda^p x)^{s}}{s!}.
\]
It is clear that we can consider the constant $b$ as a summand corresponding to $s=0$. Switching in notation back from ${\bf n}_{s}$ to ${\bf m}_{s}$, this yields f\/inally the result
\[
f_0(x)=e^{\lambda^{p+1}x}\sum_{s=0}^{n}\left\{\sum_{|{\bf m}_{s}|\leq n; m_0=0}C_{n+1-|{\bf m}_{s}|}\lambda^{s-|{\bf m}_{s}|}\prod_{j=1}^{s}\binom{p+1}{m_j}\right\}\frac{(\lambda^p x)^{s}}{s!}.
\]
This is indeed the form asserted in \eqref{compo} for $f_0(x)$ (note that in the formula there one has -- due to $k=0$ -- that $m_0=0$ holds true).

Now, let us f\/ind $f_k(x)$ with $k\geq1$. By an induction on $k$ using \eqref{eqsf0} we get that
\[
f_k(x)=\frac{\lambda^k}{[k]_q!}f_0(x)+\sum_{i=0}^{k-1}\frac{[i]_q!\lambda^{k-1-i}}{[k]_q!}g_i(x).
\]
Substituting the expression of $g_i(x)$ shows that
\begin{gather*}
e^{-\lambda^{p+1}x}f_k(x)=\frac{\lambda^k}{[k]_q!}\sum_{s=0}^{n}\left\{\sum_{|{\bf m}_{s}|\leq n; m_0=0}C_{n+1-|{\bf m}_{s}|}\lambda^{s-|{\bf m}_{s}|}\prod_{j=1}^{s}\binom{p+1}{m_j}\right\}\frac{(\lambda^p x)^{s}}{s!}\\
\phantom{e^{-\lambda^{p+1}x}f_k(x)=}{} +\sum_{i=0}^{k-1}\frac{\lambda^{k-1}}{[k]_q!}
\sum_{s=0}^{n-1}\left\{\sum_{|{\bf m}_s|\leq n-1}C_{n-|{\bf m}_s|}\lambda^{s-|{\bf m}_s|}\binom{i}{m_0}\prod_{j=1}^s\binom{p+1}{m_j}\right\}\frac{(\lambda^p x)^s}{s!}\\
\phantom{e^{-\lambda^{p+1}x}f_k(x)}{} = \frac{\lambda^k}{[k]_q!}\sum_{s=0}^{n}\left\{\sum_{|{\bf m}_{s}|\leq n; m_0=0}C_{n+1-|{\bf m}_{s}|}\lambda^{s-|{\bf m}_{s}|}\binom{k}{m_0}\prod_{j=1}^{s}\binom{p+1}{m_j}\right\}\frac{(\lambda^p x)^{s}}{s!}\\
\phantom{e^{-\lambda^{p+1}x}f_k(x)=}{}+\frac{\lambda^{k}}{[k]_q!}
\sum_{s=0}^{n-1}\left\{\sum_{|{\bf m}_s|\leq n-1}C_{n-|{\bf m}_s|}\lambda^{s-|{\bf m}_s|-1}\binom{k}{m_0+1}\prod_{j=1}^s\binom{p+1}{m_j}\right\}\frac{(\lambda^p x)^s}{s!}\\
\phantom{e^{-\lambda^{p+1}x}f_k(x)}{} =\frac{\lambda^k}{[k]_q!}\sum_{s=0}^{n}\left\{\sum_{|{\bf m}_{s}|\leq n; m_0=0}C_{n+1-|{\bf m}_{s}|}\lambda^{s-|{\bf m}_{s}|}\binom{k}{m_0}\prod_{j=1}^{s}\binom{p+1}{m_j}\right\}\frac{(\lambda^p x)^{s}}{s!}\\
\phantom{e^{-\lambda^{p+1}x}f_k(x)=}{}+\frac{\lambda^{k}}{[k]_q!}
\sum_{s=0}^{n-1}\left\{\sum_{|{\bf m}_s|\leq n; m_0>0}C_{n+1-|{\bf m}_s|}\lambda^{s-|{\bf m}_s|}\binom{k}{m_0}\prod_{j=1}^s\binom{p+1}{m_j}\right\}\frac{(\lambda^p x)^s}{s!}.
\end{gather*}
It follows that
\[
f_k(x)=e^{\lambda^{p+1}x}\frac{\lambda^k}{[k]_q!}\sum_{s=0}^{n}\left\{\sum_{|{\bf m}_{s}|\leq n}C_{n+1-|{\bf m}_{s}|}\lambda^{s-|{\bf m}_{s}|}\binom{k}{m_0}\prod_{j=1}^{s}\binom{p+1}{m_j}\right\}\frac{(\lambda^p x)^{s}}{s!},
\]
which completes the proof.
\end{proof}

Using the expression given in \eqref{compo} for $f_k(x)$ (see Lemma~\ref{lemmacompo}) together with $f(x,\theta)=\sum_{k=0}^pf_k(x)\theta^k$, we obtain the following result.

\begin{theorem}\label{Jordan}
Let $f(x,\theta)=\sum_{k=0}^pf_k(x)\theta^k$ be a solution of the differential equation
\[
({\mathcal{D}}-\lambda)^nf(x,\theta)=0,\qquad 1 \leq n\leq p.
\] Then there exist $C_1,\ldots,C_n\in\mathbb{C}$ such that
\[
f(x,\theta)=e^{\lambda^{p+1}x}\sum_{k=0}^{p}\left\{\sum_{s=0}^{n-1}\left(\sum_{|{\bf m}_s|\leq n-1}C_{n-|{\bf m}_s|}\lambda^{s-|{\bf m}_s|}\binom{k}{m_0}\prod_{i=1}^s\binom{p+1}{m_i}\right)\frac{(\lambda^p x)^s}{s!}\right\} \frac{(\lambda\theta)^k}{[k]_q!},
\]
where ${\bf m}_s=(m_0,m_1,\dots,m_s)$ with $m_0\geq0$ and $m_i\geq1$, for all $i=1,2,\ldots,s$.
\end{theorem}

\begin{example} Let us consider the case $n=2$ (the case $n=1$ was already considered above during the induction). The formula given in Theorem~\ref{Jordan} yields for $s$ the two summands corresponding to $s=0$ and $s=1$. Let us consider f\/irst $s=0$. It follows that $ |{\bf m}_0| \leq 1$ and, therefore, that $m_0=0$ or $m_0=1$. The entire inner sum yields thus $C_2+C_1\lambda^{-1}k$. In the case $s=1$ the only possibility is ${\bf m}_1=(m_0,m_1)=(0,1)$ with $|{\bf m}_1|=1$. The inner sum yields in this case $C_1 (p+1)\lambda^px$. Thus,
\begin{gather*}
f(x,\theta) = e^{\lambda^{p+1}x}\sum_{k=0}^{p}\left\{C_2+C_1\lambda^{-1}k+C_1 (p+1)\lambda^px\right\} \frac{(\lambda\theta)^k}{[k]_q!}\\
\phantom{f(x,\theta)}{}  =  C_2e^{\lambda^{p+1}x}\sum_{k=0}^{p}\frac{(\lambda\theta)^k}{[k]_q!}+ C_1\left((p+1)\lambda^px e^{\lambda^{p+1}x}\sum_{k=0}^{p}\frac{(\lambda\theta)^k}{[k]_q!}+  e^{\lambda^{p+1}x}\sum_{k=0}^{p}\frac{k\lambda^{k-1}\theta^k}{[k]_q!}\right)\\
\phantom{f(x,\theta)}{} =  C_2 e_q(\lambda^{p+1}x;\lambda\theta)+C_1\left((p+1)\lambda^p x e_q(\lambda^{p+1}x;\lambda\theta)+e^{\lambda^{p+1}x}\sum_{k=1}^{p}\frac{k\lambda^{k-1}\theta^k}{[k]_q!}\right).
\end{gather*}
This is exactly equation~\eqref{eqsecond} given in Theorem~\ref{second} for the case $n=2$ (with degenerated eigenvalue).
\end{example}

Using the linearity of the general case, equation~\eqref{deggen} and Theorem~\ref{Jordan} we get the following theorem.
\begin{theorem}\label{main} Let the differential equation \eqref{deggen} with $1\leq n \leq p$ be given. Assume that the different roots $\lambda_1,\ldots,\lambda_m$ of the characteristic polynomial $x^n+c_1x^{n-1}+\cdots+c_{n-1}x+c_n$ have multiplicities $n_1,\ldots,n_m$ $($with $n_1+\cdots + n_m =n$ and $n_i\geq 1)$. Then the solution $f(x,\theta)$ of the differential equation~\eqref{deggen} can be written as
\begin{gather*}
f(x,\theta)=\sum_{i=1}^m\! e^{\lambda_i^{p+1}x}\!\sum_{k=0}^{p}\!\left\{\sum_{s=0}^{n_i-1}\!\!\left(\sum_{|{\bf m}_s|\leq n_i-1}\!\!\!C^{(i)}_{n_i-|{\bf m}_s|}\lambda_i^{s-|{\bf m}_s|}\!\binom{k}{m_0}\!\prod_{i=1}^s\!\binom{p+1}{m_i}\!\right)\!\frac{(\lambda_i^p x)^s}{s!}\!\right\} \! \frac{(\lambda_i\theta)^k}{[k]_q!},
\end{gather*}
where $C_1^{(i)},\ldots,C^{(i)}_{n_i}\in\mathbb{C}$ are some constants $($there are $n$ of them$)$ and ${\bf m}_s=(m_0,m_1,\dots,m_s)$ with $m_0\geq0$ and $m_i\geq1$, for all $i=1,2,\ldots,s$.
\end{theorem}

This theorem immediately implies the following corollary.

\begin{corollary} Let the differential equation \eqref{deggen} with $1\leq n \leq p$ be given. The set of its solutions is a complex linear space of dimension $n$. The set of solutions of the corresponding inhomogeneous differential equation
\[
\big({\mathcal{D}}^n+c_1{\mathcal{D}}^{n-1}+ \cdots + c_{n-1}{\mathcal{D}}+ c_n\big)f(x,\theta)=g(x,\theta)
\]
is an $n$-dimensional complex affine space.
\end{corollary}
\begin{proof} It only remains to check the last assertion. The argument is the same as in the usual case: If $f_{i}(x,\theta)$ is a particular solution of the inhomogeneous dif\/ferential equation then any function of the form $f_{hom}(x,\theta)+f_{i}(x,\theta)$ -- where $f_{hom}(x,\theta)$ is a solution of the homogeneous dif\/ferential equation -- satisf\/ies the inhomogeneous dif\/ferential equation.
\end{proof}

\begin{remark} An approach similar to the case $n=2$ is also possible in the more general case where $2\leq n \leq p$. Inserting $f(x,\theta)=f_0(x)+\cdots + f_p(x)\theta^p$ into \eqref{deggen} yields after a straightforward, but slightly tedious computation in close analogy to \eqref{rec}--\eqref{recdeg2f1}
the following system
\begin{gather}\label{nrec}
f_{k+n}(x) =
-\sum_{i=1}^{n} c_i \frac{[k+n-i]_q!}{[k+n]_q!}f_{k+n-i}(x),\qquad 0\leq k \leq p-n, \\  \partial_xf_{k+n-p-1}(x) = -\sum_{i=1}^{k+n-p-1}c_i \frac{[k+n-p-1-i]_q!}{[k+n-p-1]_q!}\partial_x f_{k+n-p-1-i}(x) \nonumber\\  \phantom{\partial_xf_{k+n-p-1}(x) =}{}   -\sum_{i=k+n-p}^{n}c_i \frac{[k+n-i]_q!}{[k+n-p-1]_q!}f_{k+n-i}(x),\qquad p-n+1 \leq k\leq p.\label{nrecdeg}
\end{gather}
Note that this system constitutes indeed $p+1$ conditions as it should and that this system reduces for $n=2$ to the one given in \eqref{rec}--\eqref{recdeg2f1} (here the recursion relations \eqref{nrec} are also vacuous in the case $p=1$ and only the system of dif\/ferential equations remains -- as in the case $n=2$). Considering the recursion relation and abbreviating
$\gamma_i(k):=-c_i\frac{[k+n-i]_q!}{[k+n]_q!}$, we may write~\eqref{nrec} as
\[
\left(\begin{array}{c} f_{k+1}(x)\\ \vdots \\ f_{k+n}(x)\end{array}
\right) = \left(\begin{array}{cccccc} 0 & 1& 0 & \cdots & \cdots & 0\\
\vdots &\ddots &1&\ddots&& \vdots \\ \vdots &&\ddots & \ddots&\ddots& \vdots \\ \vdots
&&&\ddots &1&0 \\ 0 &\cdots&\cdots&\cdots&0&1\\ \gamma_n(k)&  \gamma_{n-1}(k)& \cdots & & \cdots
&  \gamma_1(k)\end{array}
\right)\left(\begin{array}{c} f_{k}(x)\\ \vdots \\ f_{k+n-1}(x)\end{array}
\right).
\]
Let us denote the above $n\times n$ matrix by $\TT_{k+1}^{(n)}$ and the
vector on the right hand side by $v_k(x)$; observe that $\TT_{k+1}^{(2)}$
equals $\TT_{k+1}$ from \eqref{notat}. We may, therefore, write the
above equation as $v_{k+1}(x)=\TT_{k+1}^{(n)}v_k(x)$. This implies that
$v_{p-n}(x)=\TT_{p-n}^{(n)}\cdots\TT_{1}^{(n)}v_0(x)$. Let
$\FF^{(n)}_{k+1}:=\TT_{k+1}^{(n)}\cdots\TT_1^{(n)}$. As in the case $n=2$ considered in the last section, we may reformulate the recursion $\FF^{(n)}_{k+1}=\TT_{k+1}^{(n)}\FF^{(n)}_k$ by considering
appropriately normalized entries of the matrices $\FF^{(n)}_k$. More
precisely, we introduce in analogy to
(\ref{2mat}) the entries $f^{(j)}_m$ by
\[
\FF^{(n)}_k=\left(\frac{f_{k-n+i}^{(j)}}{[k-n+i+1]_q!}\right)_{i,j=1,\ldots, n}.
\]
This implies for each $j$ with $j=1,\ldots,n$ the recursion relation
\begin{gather}\label{fibgen}
f_{k+n}^{(j)}=-c_1f_{k+n-1}^{(j)}-c_2f_{k+n-2}^{(j)}-\cdots-c_nf_{k}^{(j)}.
\end{gather}
Thus, for each $j$ there will be an $n$-generalized Fibonacci sequence
$(f_k^{(j)})_{k\in \mathbb{N}}$ satisfying
(\ref{fibgen}) \cite{miles,lev}; they are dif\/fering only in the initial values
$(f_0^{(j)},\ldots,f_{n-1}^{(j)})$. The initial values of
$(f_k^{(j)})_{k\in \mathbb{N}}$, are exactly the normalized entries in
the $j$-th column of $\TT_1^{(n)}$. This is the generalization of~\eqref{fibo}. To proceed further in close analogy to the case $n=2$ one denotes the zeroes of the generating function $x^n+c_1x^{n-1}+\cdots + c_n$ by $\lambda_1,\ldots,\lambda_n$, where some of these zeroes may occur with multiplicity greater than one. By using the appropriate generalization of Binet's formula (cf.~\cite{miles,lev,lee2}) one can write the general solution as $f_k=a_1\lambda_1^k+\cdots + a_n \lambda_n^k$, where the $a_i$ are determined by the initial values. However, the resulting expressions become rather unpleasant so that we won't consider this procedure further. It is, however, clear that this will lead in the second step~-- after inserting the resulting relations into~\eqref{nrecdeg}~-- to a system of linear dif\/ferential equations which has then to be solved.
\end{remark}

In the above considerations concerning \eqref{deggen} we have always assumed that $2\leq n \leq p$. Recall that the motivation for the restriction $n\leq p$ comes from the periodicity \eqref{rootp} which shows that in this case derivatives of at most f\/irst order in the components $f_k(x)$ appear, see \eqref{covderi}, whereas in the case $n\geq p+1$ derivatives of higher order appear, see \eqref{genabl}. In the remaining part of the section we consider the case where an arbitrary $n$ is allowed. It is to be expected that \eqref{deggen} is equivalent to a system of ordinary dif\/ferential equations of higher order.

\begin{proposition}\label{propgen} Let the differential equation \eqref{deggen} be given where $n=n'(p+1)+n''$ with $0\leq n'' < p+1$. This differential equation for $f(x,\theta)=f_0(x)+f_1(x)\theta+\cdots+f_p(x)\theta^p$ is equivalent to the following system of differential equations for the components $f_k(x)$
\begin{gather*}
\sum_{\nu=0}^{n''}c_{n''-\nu}[\nu]_q!\partial_x^{n'} f_{\nu}(x) = -\sum_{\mu=0}^{n'-1}\sum_{\nu=0}^{p} c_{(n'-\mu)(p+1)+(n''-\nu)}[\nu]_q!\partial_x^{\mu} f_{\nu}(x),\\ \sum_{\nu=0}^{n''}c_{n''-\nu}[l+\nu]_q!\partial_x^{n'} f_{l+\nu}(x) = -\sum_{\mu=0}^{n'-1}\bigg(\sum_{\nu=0}^{p-l} c_{(n'-\mu)(p+1)+(n''-\nu)}[l+\nu]_q!\partial_x^{\mu} f_{l+\nu}(x)\\
\phantom{\sum_{\nu=0}^{n''}c_{n''-\nu}[l+\nu]_q!\partial_x^{n'} f_{l+\nu}(x) =}{}
+\sum_{\nu=p-l+1}^{p}c_{(n'-\mu)(p+1)+(n''-\nu)}[\underline{l+\nu}]_q!\partial_x^{\mu} f_{\underline{l+\nu}}(x)\bigg), \\
\sum_{\nu=p+1-l}^{n''}c_{n''-\nu}[\underline{l+\nu}]_q!\partial_x^{n'+1} f_{\underline{l+\nu}}(x)  =  - \sum_{\nu=0}^{p-l}c_{n''-\nu}[l+\nu]_q!\partial_x^{n'} f_{l+\nu}(x)\\
 \phantom{\sum_{\nu=p+1-l}^{n''}c_{n''-\nu}[\underline{l+\nu}]_q!\partial_x^{n'+1} f_{\underline{l+\nu}}(x)  =}{}
 -\sum_{\mu=0}^{n'-1}\Bigg(\sum_{\nu=0}^{p-l} c_{(n'-\mu)(p+1)+(n''-\nu)}[l+\nu]_q!\partial_x^{\mu} f_{l+\nu}(x)\\ \phantom{\sum_{\nu=p+1-l}^{n''}c_{n''-\nu}[\underline{l+\nu}]_q!\partial_x^{n'+1} f_{\underline{l+\nu}}(x)  =}{}
 +\sum_{\nu=p-l+1}^{p}c_{(n'-\mu)(p+1)+(n''-\nu)}[\underline{l+\nu}]_q!\partial_x^{\mu} f_{\underline{l+\nu}}(x)\Bigg),
\end{gather*}
where we have introduced the abbreviation $\underline{x}:=x-(p+1)$ as well as $c_0:=1$. The second equation holds for all $l$ with $1\leq l \leq p-n''$ and the third equation for all $l$ with $p-n''+1 \leq l \leq p$.
\end{proposition}
Before turning to the proof we would like to point out a few facts. Note f\/irst that there are in total $p+1$ equations which have to be satisf\/ied, as it should be. As a second point note that there are $n''$ dif\/ferential equations of order $n'+1$ as well as $p-n''+1$ dif\/ferential equations of order~$n'$. The case $n\leq p$ considered before corresponds to $n'=0$ and $n''=n$: The above proposition yields that there are $n$ dif\/ferential equations of order $1$ as well as $p-n+1$ dif\/ferential equations of order $0$ -- which is indeed the case, see \eqref{nrec}--\eqref{nrecdeg}. On the other hand, if $n=n'(p+1)$, i.e., $n''=0$, then one has $p+1$ dif\/ferential equations of order $n'$ (and none of higher order). Thus, in this case all dif\/ferential equations have the same order.

\begin{corollary}\label{corospace} Let the differential equation \eqref{deggen} be given where $n=n'(p+1)+n''$ with $0\leq n'' < p+1$. Then  \eqref{deggen} is equivalent to a set of $n''$ differential equations of order $n'+1$ as well as $p-n''+1$ differential equations of order $n'$ for the components $f_k(x)$. In particular, the set of solutions of \eqref{deggen} is a complex linear space of dimension $n$.
\end{corollary}

\begin{proof} %(Corollary~\ref{corospace})
It remains to check the dimension of the space of solutions. However, since the linear dif\/ferential equation of order $n'$ (resp. $n'+1$) has $n'$ (resp. $n'+1$) linear independent solutions, one has in total $n''(n'+1)+(p-n''+1)n'=n'(p+1)+n''=n$ linear independent solutions.
\end{proof}

Let us check that the above equations given in Proposition~\ref{propgen} do indeed reduce to the ones given in \eqref{nrec}--\eqref{nrecdeg} for the case $n\leq p$, i.e., $n'=0$ and $n''=n$. Consider the f\/irst equation of Proposition~\ref{propgen}. Since $n'=0$ the right-hand side vanishes. It follows that $c_0 [n]_q! f_{n}(x)=-\sum_{\nu=0}^{n-1}c_{n-\nu}[\nu]_q! f_{\nu}(x)$. Recalling $c_0=1$ and introducing $i=n-\nu$, this is equivalent to
\[
f_{n}(x)=-\sum_{\nu=0}^{n-1}c_{i}\frac{[n-i]_q!}{[n]_q!} f_{n-i}(x)
\]
which is the case $k=0$ of \eqref{nrec}. In the second equation of Proposition~\ref{propgen} the right-hand side also vanishes, implying $\sum_{\nu=0}^{n}c_{n-\nu}[l+\nu]_q! f_{l+\nu}(x)=0$. Using $c_0=1$ and introducing $i=n-\nu$ yields as above
\[
 f_{l+n}(x)=-\sum_{i=1}^{n}c_{i}\frac{[l+n-i]_q!}{[l+n]_q!} f_{l+n-i}(x)
\]
which are the (remaining) cases $1\leq l\leq p-n$ of \eqref{nrec}. In the third equation of Proposition~\ref{propgen} only the f\/irst sum remains, implying
\[
\sum_{\nu=p+1-l}^{n}c_{n-\nu}[l+\nu-(p+1)]_q!\partial_x f_{l+\nu-(p+1)}(x) = - \sum_{\nu=0}^{p-l}c_{n-\nu}[l+\nu]_q! f_{l+\nu}(x).
\]
Singling out again the summand $\nu=n$, recalling $c_0=1$ and introducing $i=n-\nu$ yields
\begin{gather*}
[l+n-(p+1)]_q!\partial_x f_{l+n-(p+1)}(x) =  -\sum_{i=1}^{l+n-(p+1)}c_{i}[l+n-(p+1)-i]_q!\partial_x f_{l+n-(p+1)-i}(x) \\
\phantom{[l+n-(p+1)]_q!\partial_x f_{l+n-(p+1)}(x) =}{} - \sum_{i=n+l-p}^{n}c_{i}[l+n-i]_q! f_{l+n-i}(x)
\end{gather*}
which is exactly \eqref{nrecdeg}. Thus, the system of equations given in Proposition~\ref{propgen} reduces for $n\leq p$ indeed to \eqref{nrec}--\eqref{nrecdeg}. Now, let us turn to the proof of the proposition.

\begin{proof}[Proof of Proposition~\ref{propgen}] The proof is straightforward but slightly tedious since one has to be very careful with the indices involved. Let us introduce $c_0:=1$ as well as $d_s:=c_{n-s}$ for $0\leq s \leq n$. Recalling furthermore that we can split $n$ in a unique fashion as $n=n'(p+1)+n''$ with $0\leq n''\leq p$, we can write the left-hand side of \eqref{deggen} in the following form
\[
\sum_{s=0}^nd_s\mathcal{D}^sf(x,\theta)=\sum_{\mu=0}^{n'-1}\sum_{\nu=0}^{p}d_{\mu(p+1)+\nu}\mathcal{D}^{\mu(p+1)+\nu}f(x,\theta)+ \sum_{\nu=0}^{n''}d_{n'(p+1)+\nu}\mathcal{D}^{n'(p+1)+\nu}f(x,\theta).
\]
Inserting the expansion $f(x,\theta)=f_0(x)+f_1(x)\theta+\cdots+f_p(x)\theta^p$ and using \eqref{genabl} yields
\begin{gather*}
\sum_{s=0}^nd_s\mathcal{D}^sf(x,\theta) =  \sum_{\mu=0}^{n'-1}\sum_{\nu=0}^{p}\sum_{k=0}^{\nu -1} d_{\mu(p+1)+\nu}\frac{[k]_q!}{[p-(\nu    -1)+k]_q!}\partial_x^{\mu +1} f_k(x)\theta^{p-(\nu -1)+k}\\
 \phantom{\sum_{s=0}^nd_s\mathcal{D}^sf(x,\theta) =}{} + \sum_{\mu=0}^{n'-1}\sum_{\nu=0}^{p}\sum_{k=\nu}^{p} d_{\mu(p+1)+\nu}\frac{[k]_q!}{[k-\nu]_q!}\partial_x^{\mu} f_k(x)\theta^{k-\nu} \\
  \phantom{\sum_{s=0}^nd_s\mathcal{D}^sf(x,\theta) =}{}
  + \sum_{\nu=0}^{n''}\sum_{k=0}^{\nu -1} d_{n'(p+1)+\nu}\frac{[k]_q!}{[p-(\nu    -1)+k]_q!}\partial_x^{n'+1} f_k(x)\theta^{p-(\nu -1)+k}\\
  \phantom{\sum_{s=0}^nd_s\mathcal{D}^sf(x,\theta) =}{}
  + \sum_{\nu=0}^{n''}\sum_{k=\nu}^{p} d_{n'(p+1)+\nu}\frac{[k]_q!}{[k-\nu]_q!}\partial_x^{n'} f_k(x)\theta^{k-\nu}.
\end{gather*}
We now consider the terms on the right-hand side of this equation separately. Let us begin with the second term. Introducing $l=k-\nu$, we switch from the indices $k$, $\nu$ to $l$, $\nu$ and write this term equivalently as
\[
\sum_{l=0}^{p}\sum_{\nu=0}^{p-l}\sum_{\mu=0}^{n'-1} d_{\mu(p+1)+\nu}\frac{[l+\nu]_q!}{[l]_q!}\partial_x^{\mu} f_{l+\nu}(x)\theta^{l}.
\]
Let us turn to the fourth term. Here we also introduce $l=k-\nu$ and switch from the indices~$k$,~$\nu$ to $l$, $\nu$. Treating the range of $\nu$ carefully, this term can equivalently be written as
\[
\sum_{l=0}^{p}\sum_{\nu=0}^{\min(p-l,n'')}d_{n'(p+1)+\nu}\frac{[l+\nu]_q!}{[l]_q!}\partial_x^{n'} f_{l+\nu}(x)\theta^{l}.
\]
In the f\/irst term we introduce $l=p+1-(\nu-k)$; it ranges from $1$ to $p$. Switching from the indices $k$, $\nu$ to $l$, $\nu$, this term is equivalent to
\[
\sum_{l=1}^{p}\sum_{\nu=p+1-l}^{p}\sum_{\mu=0}^{n'-1}d_{\mu(p+1)+\nu}\frac{[l+\nu-(p+1)]_q!}{[l]_q!}\partial_x^{\mu} f_{l+\nu-(p+1)}(x)\theta^{l}.
\]
Treating the third term in the same fashion (carefully treating the ranges) yields
\[
\sum_{l=p+1-n''}^{p}\sum_{\nu=p+1-l}^{n''}d_{n'(p+1)+\nu}\frac{[l+\nu-(p+1)]_q!}{[l]_q!}\partial_x^{n'+1} f_{l+\nu-(p+1)}(x)\theta^{l}.
\]
Inspecting the resulting expressions for the four terms we see that upon combining them in powers of $\theta$ there are three ranges to consider: 1) The range $l=0$; here one has a contribution from the second and fourth term. 2) The range $1\leq l \leq p-n''$; here one has a contribution from the second and fourth term as well as the f\/irst term. 3) The range $p-n''+1\leq l \leq p$; here all terms contribute. We can, therefore, write
\[
\sum_{s=0}^nd_s\mathcal{D}^sf(x,\theta)= E_0(x)+ \sum_{l=1}^{p-n''} F_l(x) \theta^l + \sum_{l=p+1-n''}^{p} G_l(x) \theta^l,
\]
where we have abbreviated the corresponding functions $E_0(x)$, $F_l(x)$ and $G_l(x)$ as follows:
\begin{gather*}
E_0(x) = \sum_{\nu=0}^{p}\sum_{\mu=0}^{n'-1} d_{\mu(p+1)+\nu}[\nu]_q!\partial_x^{\mu} f_{\nu}(x)+ \sum_{\nu=0}^{n''}d_{n'(p+1)+\nu}[\nu]_q!\partial_x^{n'} f_{\nu}(x), \\
F_l(x) = \sum_{\nu=0}^{p-l}\sum_{\mu=0}^{n'-1} d_{\mu(p+1)+\nu}\frac{[l+\nu]_q!}{[l]_q!}\partial_x^{\mu} f_{l+\nu}(x)+\sum_{\nu=0}^{\min(p-l,n'')}d_{n'(p+1)+\nu}\frac{[l+\nu]_q!}{[l]_q!}\partial_x^{n'} f_{l+\nu}(x)\\
\phantom{F_l(x) =}{} +\sum_{\nu=p+1-l}^{p}\sum_{\mu=0}^{n'-1}d_{\mu(p+1)+\nu}\frac{[\underline{l+\nu}]_q!}{[l]_q!}\partial_x^{\mu} f_{\underline{l+\nu}}(x), \\
G_l(x) =  F_l(x)+\sum_{\nu=p+1-l}^{n''}d_{n'(p+1)+\nu}\frac{[\underline{l+\nu}]_q!}{[l]_q!}\partial_x^{n'+1} f_{\underline{l+\nu}}(x).
\end{gather*}
Thus, \eqref{deggen} is equivalent to $E_0(x)=0$, $F_l(x)=0=G_l(x)$ for all $l$. Let us start with $E_0(x)=0$. This is equivalent to
\[
\sum_{\nu=0}^{n''}d_{n'(p+1)+\nu}[\nu]_q!\partial_x^{n'} f_{\nu}(x)=-\sum_{\mu=0}^{n'-1}\sum_{\nu=0}^{p} d_{\mu(p+1)+\nu}[\nu]_q!\partial_x^{\mu} f_{\nu}(x).
\]
Recalling $d_s=c_{n-s}$ as well as $n=n'(p+1)+n''$, this is equal to
\[
\sum_{\nu=0}^{n''}c_{n''-\nu}[\nu]_q!\partial_x^{n'} f_{\nu}(x)=-\sum_{\mu=0}^{n'-1}\sum_{\nu=0}^{p} c_{(n'-\mu)(p+1)+(n''-\nu)}[\nu]_q!\partial_x^{\mu} f_{\nu}(x),
\]
which is the f\/irst asserted equation. Let us turn to $F_l(x)=0$. The highest order of a derivative appearing is $n'$. Thus, switching also to the coef\/f\/icients $c_s$, we can write $F_l(x)=0$ (with $1\leq l \leq p-n''$) equivalently as
\begin{gather*}
\sum_{\nu=0}^{n''}c_{n''-\nu}[l+\nu]_q!\partial_x^{n'} f_{l+\nu}(x) = -\sum_{\mu=0}^{n'-1}\Bigg(\sum_{\nu=0}^{p-l} c_{(n'-\mu)(p+1)+(n''-\nu)}[l+\nu]_q!\partial_x^{\mu} f_{l+\nu}(x)\\ \phantom{\sum_{\nu=0}^{n''}c_{n''-\nu}[l+\nu]_q!\partial_x^{n'} f_{l+\nu}(x) =}{}
+\sum_{\nu=p-l+1}^{p}c_{(n'-\mu)(p+1)+(n''-\nu)}[\underline{l+\nu}]_q!\partial_x^{\mu} f_{\underline{l+\nu}}(x)\Bigg),
\end{gather*}
which is the asserted second equation. It remains to consider $G_l(x)=0$. Here derivatives of order $n'+1$ appear. Thus, switching also to the coef\/f\/icients $c_s$, we can write $G_l(x)=0$ (with $p-n''+1 \leq l \leq p$) equivalently as
\begin{gather*}
\sum_{\nu=p+1-l}^{n''}c_{n''-\nu}[\underline{l+\nu}]_q!\partial_x^{n'+1} f_{\underline{l+\nu}}(x)  =  - \sum_{\nu=0}^{p-l}c_{n''-\nu}[l+\nu]_q!\partial_x^{n'} f_{l+\nu}(x)\\
\phantom{\sum_{\nu=p+1-l}^{n''}c_{n''-\nu}[\underline{l+\nu}]_q!\partial_x^{n'+1} f_{\underline{l+\nu}}(x)  =}{}
-\sum_{\mu=0}^{n'-1}\Bigg(\sum_{\nu=0}^{p-l} c_{(n'-\mu)(p+1)+(n''-\nu)}[l+\nu]_q!\partial_x^{\mu} f_{l+\nu}(x)\\ \phantom{\sum_{\nu=p+1-l}^{n''}c_{n''-\nu}[\underline{l+\nu}]_q!\partial_x^{n'+1} f_{\underline{l+\nu}}(x)  =}{}
+\sum_{\nu=p-l+1}^{p}c_{(n'-\mu)(p+1)+(n''-\nu)}[\underline{l+\nu}]_q!\partial_x^{\mu} f_{\underline{l+\nu}}(x)\Bigg),
\end{gather*}
which is the asserted third equation.
\end{proof}

\section{Linear systems of f\/irst order}\label{section6}

Using the standard procedure to transform a linear dif\/ferential equation into a system of dif\/ferential equations of f\/irst order, we def\/ine $v_1(x,\theta):=f(x,\theta)$ as well as $v_j(x,\theta):={\mathcal{D}}^{j-1} f(x,\theta)$ for $2\leq j \leq n$. Forming the vector function $v(x,\theta):=(v_1(x,\theta),\ldots,v_{n}(x,\theta))^t$, we can transform the dif\/ferential equation \eqref{deggen} into the form ${\mathcal{D}}v(x,\theta)=Av(x,\theta)$ where the matrix $A\in \mathbb{C}^{n,n}$ is given in the usual form. Clearly, one is interested in the general system
\begin{gather}\label{system}
{\mathcal{D}}w(x,\theta)=Aw(x,\theta),
\end{gather}
where the matrix $A\in \mathbb{C}^{n,n}$ is arbitrary and $w(x,\theta)=(w_1(x,\theta),\ldots,w_{n}(x,\theta))^t$ is a function to be determined. What one is in fact looking for is the generalization of the function $e_q(a^{p+1}x;a\theta)\equiv e^{a^{p+1}x}\sum_{k=0}^p\frac{(a\theta)^k}{[k]_q!}$ satisfying ${\mathcal{D}}e_q(a^{p+1}x;a\theta)=ae_q(a^{p+1}x;a\theta)$ to the case with matrix arguments.

\begin{proposition}\label{propmexp} Let $A\in \mathbb{C}^{n,n}$ be given. For a given $p\geq 1$ define the matrix exponential function~$E_q$ by
\[
E_q(A^{p+1}x;A\theta):=e^{A^{p+1}x}\sum_{k=0}^p \frac{(A\theta)^k}{[k]_q!} .
\]
The so defined function $E_q$ has the following properties:
\begin{enumerate}\itemsep=0pt
\item[$1.$] In the scalar case $n=1$ it reduces to the exponential function $e_q$.
\item[$2.$] If the matrix $A$ has diagonal form, i.e., $A=\mbox{\rm diag}(a_1,\ldots,a_n)$, then one has
\[
E_q(A^{p+1}x;A\theta)=\mbox{\rm diag}\left\{e_q(a_1^{p+1}x;a_1\theta),\ldots,e_q(a_n^{p+1}x;a_n\theta)\right\}.
\]
\item[$3.$] It satisfies ${\mathcal{D}}E_q(A^{p+1}x;A\theta)=AE_q(A^{p+1}x;A\theta)$.
\item[$4.$] It can also be written as $E_q(A^{p+1}x;A\theta)=\sum_{k=0}^p \frac{(A\theta)^k}{[k]_q!}e^{A^{p+1}x} $.
\end{enumerate}
\end{proposition}

\begin{proof} The proof of the f\/irst property is clear. The second property is shown by a straightforward computation and the third property follows as in the scalar case by a direct calculation from the def\/initions. The last property follows since powers of $A$ commute.
\end{proof}

Let us now consider the general linear system \eqref{system}. Recalling the third property of $E_q$ stated in Proposition \ref{propmexp}, it is clear that $w(x,\theta):=E_q(A^{p+1}x;A\theta)c$ - where $c=(c_1,\ldots,c_n)^t$ with $c_i\in \mathbb{C}$ is a vector of constants -- is a solution of~\eqref{system}. It remains to be shown that no other solutions exist. This will be done by an explicit construction of the solution.
\begin{theorem} Let the linear system~\eqref{system} with $A\in \mathbb{C}^{n,n}$ be given. Any solution $w(x,\theta)$ of~\eqref{system} can be written as
\[
w(x,\theta)=E_q(A^{p+1}x;A\theta)c,
\]
where $c=(c_1,\ldots,c_n)^t$ with $c_i\in \mathbb{C}$ is a vector of constants. In particular, the set of solutions is a complex linear space of dimension $n$.
\end{theorem}
\begin{proof} To show that all solutions can be written in this fashion we prove the asserted formula by constructing the solution. Inserting the function $w(x,\theta)=(w_1(x,\theta),\ldots,w_n(x,\theta))^t$ into \eqref{system} yields the following system
\begin{gather*}
\frac{1}{[p]_q!}\partial_x w_{i,0}(x) = \sum_{l=1}^n a_{il}w_{l,p}(x),\qquad 1 \leq i \leq n, \\
\left[k\right]_q w_{i,k}(x) =  \sum_{l=1}^n a_{il}w_{l,k-1}(x),\qquad 1 \leq i \leq n,\qquad 1 \leq k \leq p,
\end{gather*}
which is equivalent to
\begin{gather*}
\left(\begin{array}{c} \partial_x w_{1,0}(x)\\ \vdots \\ \partial_x w_{n,0}(x)\end{array}\right)= [p]_q! A\left(\begin{array}{c} w_{1,p}(x)\\ \vdots \\ w_{n,p}(x)\end{array}\right),\\
 \left(\begin{array}{c} w_{1,k}(x)\\ \vdots \\ w_{n,k}(x)\end{array}\right)= \frac{A}{[k]_q} \left(\begin{array}{c} w_{1,k-1}(x)\\ \vdots \\ w_{n,k-1}(x)\end{array}\right), \qquad 1\leq k \leq p.
\end{gather*}
A simple iteration shows that
\[
\left(\begin{array}{c} w_{1,k}(x)\\ \vdots \\ w_{n,k}(x)\end{array}\right)= \frac{A^k}{[k]_q!} \left(\begin{array}{c} w_{1,0}(x)\\ \vdots \\ w_{n,0}(x)\end{array}\right), \qquad 1\leq k \leq p,
\]
and, consequently,
\[
\left(\begin{array}{c} \partial_x w_{1,0}(x)\\ \vdots \\ \partial_x w_{n,0}(x)\end{array}\right)=A^{p+1}\left(\begin{array}{c} w_{1,0}(x)\\ \vdots \\  w_{n,0}(x)\end{array}\right).
\]
Thus, there exists a vector $c=(c_1,\ldots,c_n)^t$ of constants $c_i\in \mathbb{C}$ such that
\[
\left(\begin{array}{c} w_{1,0}(x)\\ \vdots \\ w_{n,0}(x)\end{array}\right)=e^{A^{p+1}x}\left(\begin{array}{c} c_1\\ \vdots \\  c_n \end{array}\right)=e^{A^{p+1}x}c.
\]
Collecting the above results shows that
\begin{gather*}
w(x,\theta) =  \left(\begin{array}{c} \sum_{k=0}^pw_{1,k}(x)\theta^k \\ \vdots \\ \sum_{k=0}^pw_{n,k}(x)\theta^k \end{array}\right)  = \sum_{k=0}^p \left(\begin{array}{c} w_{1,k}(x)\\ \vdots \\ w_{n,k}(x)\end{array}\right)\theta^k = \sum_{k=0}^p \frac{(A\theta)^k}{[k]_q!}\left(\begin{array}{c} w_{1,0}(x)\\ \vdots \\ w_{n,0}(x)\end{array}\right)\\
 \phantom{w(x,\theta)}{} = \sum_{k=0}^p \frac{(A\theta)^k}{[k]_q!}e^{A^{p+1}x}c    =  E_q(A^{p+1}x;A\theta)c,
\end{gather*}
as asserted.
\end{proof}

\begin{remark} It is interesting to note that there is no simplif\/ication of the proof or the resulting formula for the case $p=1$ -- in contrast to the scalar case where some simplif\/ications occur.
\end{remark}

\section{Some remarks concerning variable coef\/f\/icients}\label{section7}

It is very important to note that we have here two types of variables, namely the ordinary ``bosonic'' one, i.e., $x$, and a nilpotent ``fermionic'' one, i.e., $\theta$. The fermionic variable is more an algebraic object then an analytical one, but here the two variables are mixed. Let us consider the linear dif\/ferential equation with variable coef\/f\/icients, i.e., an equation of the form
\[
({\mathcal{D}}^n+c_1(x,\theta){\mathcal{D}}^{n-1}+ \cdots + c_{n-1}(x,\theta){\mathcal{D}}+c_n(x,\theta))f(x,\theta)=g(x,\theta),
\]
where the coef\/f\/icients $c_i(c,\theta)$ are functions. We expect that it will be rather dif\/f\/icult to f\/ind solutions in general since the equation is more complex then the ``ordinary'' dif\/ferential equation involving only the bosonic variable $x$. However, if we consider instead the coef\/f\/icient functions to be depending only on the fermionic variable, i.e., $c_i(x,\theta)\equiv c_i(\theta)$, then the resulting dif\/ferential equation
\[
({\mathcal{D}}^n+c_1(\theta){\mathcal{D}}^{n-1}+ \cdots +c_{n-1}(\theta){\mathcal{D}}+ c_n(\theta))f(x,\theta)=g(x,\theta)
\]
becomes more tractable since multiplication with the functions $c_i(\theta)$ results only in a swirling of the coef\/f\/icients -- and no ``ordinary'' dif\/ferential equations with variable coef\/f\/icients have to be solved as subproblem. Instead of discussing these equations further we illustrate them with one of the most simple examples.

\begin{example} Let the dif\/ferential equation
\begin{gather}\label{varcoeff}
(\mathcal{D}-c(\theta))f(x,\theta)=0
\end{gather}
be given. Clearly, if $c(\theta)$ were constant, i.e., $c(\theta)=c$, then we would have to solve the eigenvalue equation $\mathcal{D}f(x,\theta)=cf(x,\theta)$ whose solution we have already found to be $f(x,\theta)=Ce_q(c^{p+1}x;c\theta)$. Considering \eqref{varcoeff} and writing $c(\theta){=}\sum_{l=0}^pc_l\theta^l$, the ansatz $f(x,\theta){=}\sum_{k=0}^pf_k(x)\theta^k$ yields the system
\[
\frac{1}{[p]_q!}\partial_xf_0(x)=\sum_{l=0}^pc_{p-l}f_l(x),\qquad [k+1]_q f_{k+1}(x)=\sum_{l=0}^k c_{k-l}f_l(x),\qquad 0\leq k \leq p-1.
\]
The strategy to solve this equation is clear: Iteration of the second equation yields a relation of the form $f_k(x)=C_k(c(\theta),p)f_0(x)$ which can be inserted into the f\/irst equation to obtain a dif\/ferential equation $\partial_xf_0(x)=D(c(\theta),p)f_0(x)$ which can be solved as $f_0(x)=Ce^{D(c(\theta),p)x}$. Inserting this into the result for $f_k(x)$ yields then $f_k(x)$ and, consequently, $f(x,\theta)$. However, to determine the concrete values for the coef\/f\/icients $C_k(c(\theta),p)$ and $D(c(\theta),p)$ seems to be dif\/f\/icult in the general case. As an illustration we consider the special case $p=2$. Here one obtains for $k=0$ the relation $f_1(x)=c_0f_0(x)$ and for $k=1$ the relation $[2]_q f_2(x)=c_1f_0(x)+c_0f_1(x)$, implying $f_2(x)=\frac{c_1+c_0^2}{[2]_q}f_0(x)$. These results can be inserted into the f\/irst equation and give
\[
\partial_xf_0(x)=\left\{c_0^3+(1+[2]_q)c_0c_1+[2]_qc_2 \right\}f_0(x) \equiv D(c(\theta),2)f_0(x).
\]
Thus, $f_0(x)=Ce^{D(c(\theta),2)x}$. Inserting this into the equations for the components $f_k(x)$, we have f\/inally found
\[
f(x,\theta)=Ce^{D(c(\theta),2)x}\left\{1+c_0\theta + \frac{c_0^2+c_1}{[2]_q!}\theta^2 \right\}.
\]
Note that in the case where $c(\theta)$ is constant, i.e., $c(\theta)=c$, one has $D(c,2)=c^3=c^{2+1}$ and the solution given above simplif\/ies to $f(x,\theta)=Ce^{c^{2+1}x}(1+c\theta + \frac{(c\theta)^2}{[2]_q!})=Ce_q(c^{2+1}x;c\theta)$, as it should.
\end{example}

\begin{remark} As noted already in the introduction, a certain dif\/ferential equation involving a para-Grassmann variable was solved in \cite{Alv} (by writing $f(x,\theta)=\sum_{k=0}^p f_k(x)\theta^k$, inserting this into the dif\/ferential equation and comparing coef\/f\/icients of $\theta^k$). Adapting the notation to the one used in the present paper, the dif\/ferential equation considered in \cite{Alv} can be written as
\[
\left[\partial_x + (\mu x +\nu)\sum_{l=0}^{p}\frac{(-\theta)^l}{l!}\partial^l_x\right]f(x,\theta)=\lambda f(x,\theta),
\]
where $\lambda,\mu,\nu\in \mathbb{C}$. Clearly, the structure of this dif\/ferential equation dif\/fers from the structure of the dif\/ferential equations considered in the present paper -- see, e.g., equation \eqref{easy} for a simple eigenvalue problem in our context.
\end{remark}

\section{A simple family of nonlinear dif\/ferential equations}\label{section8}

In the above sections we have considered certain types of linear
dif\/ferential equations and have found them to be very similar to
``ordinary'' dif\/ferential equations (with only ``bosonic''
variables). In this section we want to discuss some equations which
are closer to partial dif\/ferential equations in ``bosonic''
variables. Let us start with the very simple nonlinear equation $
({\mathcal{D}}f(x,\theta))^2=0$. Solving this equation is
straightforward: All one has to do is to insert \eqref{paracov} and
compare coef\/f\/icients. The result is that the general solution is given in the case $p=1,2$ by
$f(x,\theta)=f_0(x)$ and in the general case of $p \geq 3$ by
$f(x,\theta)=f_0(x)+\sum_{k=2+\left[\frac{p}{2}\right]}^p
f_k(x)\theta^k$, where we have denoted by $[x]$ the largest integer smaller or equal than
$x$. Thus, the general solution depends on
$p-\left[\frac{p}{2}\right]$ functions. More generally, one may
consider $({\mathcal{D}}^mf(x,\theta))^n=0$ for arbitrary $n\in
\mathbb{N} $ and $m\leq p$. Writing $
{\mathcal{D}}^mf(x,\theta)\equiv\sum_{l=0}^p h_l(x)\theta^l$ and
considering $(\sum_{l=0}^p h_l(x)\theta^l)^n=0$ implies the vanishing
of $h_0(x), \ldots, h_{\left[\frac{p}{n}\right]}(x)$. Applying
\eqref{covderi} shows that $h_0(x)=f_m(x)$ and that the other $h_j(x)$
are given in a similar fashion where an interesting connection
between $m$, $n$ and~$p$ appears. These results are collected in the
following theorem.
\begin{theorem} For arbitrary $p,n\in \mathbb{N}$ and $ m\leq p$ the
general solution of the differential equation $({\mathcal{D}}^m
f(x,\theta))^n=0$ is given in the case $1\leq m \leq p - \left[\frac{p}{n}\right]$ by
\[
f(x,\theta)=\sum_{k=0}^{m-1}f_k(x)\theta^k + \sum_{k=m+\left[\frac{p}{n}\right]+1}^{p}f_k(x)\theta^k
\]
and in the case
$ p -\left[\frac{p}{n}\right]<m\leq p$ by
\[
f(x,\theta)=\sum_{k=0}^{m-(p-\left[\frac{p}{n}\right])-1}C_k\theta^k +
\sum_{k=m-(p-\left[\frac{p}{n}\right])}^{m-1}f_k(x)\theta^k,
\]
where the appearing functions $f_k(x)$ are arbitrary $($note that there
are in both cases $p-\left[\frac{p}{n}\right]$ of them$)$ and the $C_k \in \mathbb{C}$
are constants. Thus, the set of solutions has infinite dimensions.
\end{theorem}

Using the ``periodicity'' described in \eqref{rootp},
the case where $m\geq p+1$ can be reduced to the one considered in the above theorem. In
fact, writing $m=m'(p+1)+m''$ with $0\leq m'' \leq p$, one has
${\mathcal{D}}^mf(x,\theta)={\mathcal{D}}^{m''}h(x,\theta)$ with $h(x,\theta)=\partial_x^{m'}f(x,\theta)$.

\section{Conclusions}\label{section9}

\looseness=1
In this paper we have considered linear dif\/ferential equations
involving a para-Grassmann variable vanishing in $(p+1)$-th order. Suppose that we are given
an equation of order $n=n'(p+1)+n''$. Expressing the parasuperfunction $f(x,\theta)$
through the components $f_k(x)$ and inserting this into the dif\/ferential equation
results in a system of $n''$ dif\/ferential equations of order $n'+1$ as well as $p-n''+1$ dif\/ferential equations of order $n'$ for the components $f_k(x)$. In the case $n\leq p$ one has, therefore, a system of $p-n+1$ recursion relations and $n$ dif\/ferential equations of f\/irst order. The key to the solution of this system consists in expressing the ``higher components'' through the ``lowest components'' and then solving the resulting system of ordinary dif\/ferential equations for the ``lowest components''. Having found the explicit form of the ``lowest components'' then allows one to describe the ``higher components'' and, consequently, $f(x,\theta)$ itself. For the linear dif\/ferential equation with constant coef\/f\/icients of order $n\leq p$ we have determined its solutions and found that the set of solutions is a complex linear space of dimension~$n$ (the latter result also holds for arbitrary $n$). Characteristic for these linear dif\/ferential equations with constant coef\/f\/icients of order $n\leq p$ is an interesting blend of ``ordinary'' dif\/ferential equations (involving only the ``bosonic'' variable $x$) and recursion relations. These recursion relations are intimately connected to the Fibonacci numbers and their generaliza\-tions to higher order. It is interesting to note that in the results there is no dif\/ference between the cases $p=1$ (i.e., an ``ordinary'' Grassmann variable) and $p>1$, although in the calculations (and proofs) there are striking dif\/ferences: In the case $p=1$ the above-mentioned recursion relations are empty (i.e., not existent) and only a system of dif\/ferential equations has to be solved. It also deserves to be mentioned that the appropriate exponential function is of utmost importance for describing the solutions. This is similar to the ``bosonic'' case but there exists a decisive dif\/ference: Here the ``derivative'' $\mathcal{D}$ is not a derivation, i.e., one does not have the usual law for the derivative of a product at hand. In particular, in the case corresponding to degenerated eigenvalues it is a~priori not clear what the analogue of the usual ansatz $x^n e^{\lambda x}$ should be in the para-Grassmann case and one of the main dif\/f\/iculties met in the present work was to solve also these cases (even with hindsight~-- i.e., the formulas for the solution at hand~-- it is not clear whether one would have made such an ansatz). Following the usual approach, the linear dif\/ferential equation with constant coef\/f\/icients can be transformed into a system of dif\/ferential equations of f\/irst order. For such a general system we determined its solution and showed in particular that the set of solutions is also a complex linear space (of expected dimension).

We also considered some other classes of dif\/ferential equations which, however, seem not as well behaved as the classes considered above and, therefore, deserve a closer study in the future. As a f\/irst example we mention the scalar linear dif\/ferential equations where variable coef\/f\/icients are allowed. We considered only an example of the case where the coef\/f\/icients are functions of the para-Grassmann variable $\theta$ (no dependency on $x$). Again, one f\/inds a blend of ``ordinary'' dif\/ferential equations and recursion relations but due to the variable coef\/f\/icients the components in the recursion relations obtain a ``twist'', resulting in slightly cumbersome expressions. It is obvious that for the general case some new ideas are needed. As a second example we considered a very simple family of nonlinear dif\/ferential equations. Due to the nilpotency of the para-Grassmann variable $\theta$ we found a behavior than dif\/fers drastically from the linear case: Although the ususal blend of recursion relations and ordinary dif\/ferential equations appears there exist (for appropriate choices of the involved parameters) several components of the solution for which no restrictions have to be satisf\/ied. Thus, the dimension of the set of solutions is in general not $m$ (as one would have expected for the dif\/ferential equation of order~$m$) but inf\/inite! Thus, the close analogy to the ordinary (``bosonic'') dif\/ferential equations found in the linear cases breaks down and one f\/inds a behaviour which reminds one more of ``bosonic'' partial dif\/ferential equations. Clearly, this is also a point where a closer study might reveal further interesting connections.

Before closing let us remark that -- in addition to the above-mentioned points concerning dif\/ferential equations involving one ``bosonic'' variable $x$ and one para-Grassmann variable~$\theta$~-- one can also consider partial dif\/ferential equations involving several bosonic and para-Grassmann variables. Since in this case the para-Grassmann variables are~-- in addition to being nilpotent~-- noncommutative, it is to be expected that it will be much harder to obtain results for these para-Grassmann partial dif\/ferential equations than for the ``ordinary'' para-Grassmann dif\/ferential equations considered in the present paper. Before beginning this journey into the unknown territory full of treasures one should prepare oneself to meet surprises~-- pleasant or otherwise~-- around every corner and it seems mandatory to become comfortable with those of them showing up already in the one-variable case. This is what we began in the present paper.

\subsection*{Acknowledgements}

The authors would like to thank V.I.~Tkach and R.M.~Yamaleev for instructive correspondence as well as the anonymous referees for several suggestions improving the paper.

\pdfbookmark[1]{References}{ref}
\LastPageEnding

\end{document}